\documentclass[
    reprint,
    superscriptaddress,
    amsmath,amssymb,
    aps,
    prb,
]{revtex4-2}

\usepackage{graphicx}
\graphicspath{{images/}}
\usepackage{dcolumn}
\usepackage{bm}
\usepackage{physics}
\usepackage{hyperref}
\usepackage{listings}
\usepackage{booktabs}
\usepackage{float}
\usepackage{xcolor}

\begin{document}

\title{STEP: Spin Tensor Equivariant Potential for Data-Efficient Learning of Magnetic Potential Energy Surfaces}

\author{Yuanqing Gao}
\affiliation{Center for Neutron Science and Technology, Guangdong Provincial Key Laboratory of Magnetoelectric Physics and Devices, State Key Laboratory of Optoelectronic Materials and Technologies, School of Physics, Sun Yat-Sen University, Guangzhou, 510275, China}

\author{Wen-Hao Luo}
\affiliation{Center for Neutron Science and Technology, Guangdong Provincial Key Laboratory of Magnetoelectric Physics and Devices, State Key Laboratory of Optoelectronic Materials and Technologies, School of Physics, Sun Yat-Sen University, Guangzhou, 510275, China}

\author{Lei Zhang}
\affiliation{Center for Neutron Science and Technology, Guangdong Provincial Key Laboratory of Magnetoelectric Physics and Devices, State Key Laboratory of Optoelectronic Materials and Technologies, School of Physics, Sun Yat-Sen University, Guangzhou, 510275, China}

\author{Kun Cao}
\email{caok7@mail.sysu.edu.cn}
\affiliation{Center for Neutron Science and Technology, Guangdong Provincial Key Laboratory of Magnetoelectric Physics and Devices, State Key Laboratory of Optoelectronic Materials and Technologies, School of Physics, Sun Yat-Sen University, Guangzhou, 510275, China}

\begin{abstract}
Accurate and efficient modeling of magnetic potential energy surfaces remains challenging because spin-polarized first-principles calculations for diverse non-collinear spin-lattice configurations are computationally demanding. Here we introduce the Spin Tensor Equivariant Potential (STEP), a magnetic machine-learning interatomic potential that treats vector magnetic moments as continuous geometric degrees of freedom and embeds them in an equivariant representation. By coupling the central spin representation to its local spin-lattice environment through a Center-Environment Tensor Product, STEP introduces a physics-informed bias while preserving translational invariance and $\mathrm{SO}(3)$ equivariance and supporting feature-level time-reversal symmetrization. Learning-curve analysis on monolayer CrI$_3$ shows that STEP achieves pronounced data efficiency, with higher-order tensor channels and iterative center-environment couplings leading to steep learning curves for energy, force, and magnetic force errors. On public FeAl, CrN, and Fe benchmarks, STEP achieves competitive or improved accuracy compared with recent magnetic machine-learning potentials. Using a compact but representative CrI$_3$ dataset, STEP reproduces phonon dispersions and magnon spectra with high fidelity, capturing subtle anisotropic magnetic interactions. For Fe$_2$Mo$_3$O$_8$, STEP further provides a quantitative description of magnon--phonon hybridization and reproduces its characteristic magnon polaron dispersion. Finally, spin dynamics simulations driven by STEP yield Curie temperatures for monolayer CrI$_3$ and bcc Fe in good agreement with experiments. These results establish STEP as a physically informed, data-efficient, and scalable framework for modeling spin-lattice coupling, magnetic excitations, and finite-temperature magnetic behavior.
\end{abstract}

\maketitle

\section{\label{sec:intro}Introduction}

Machine-learning interatomic potentials (MLIPs) have transformed atomistic modeling by approaching first-principles accuracy at a fraction of the computational cost, thereby enabling large-scale simulations of complex phenomena\cite{BP,GAP,DPMD,nequip,mace}. Extending this success to magnetic materials, however, remains particularly challenging\cite{review,magnet}. In such systems, the potential energy surface depends not only on atomic coordinates but also on local magnetic degrees of freedom. Variations in the magnetic degrees of freedom can induce extremely subtle energy changes, which are often governed by competing exchange interactions, magnetic anisotropy, and spin-orbit coupling (SOC) effects.

Recent years have witnessed several important advances in magnetic machine-learning potentials\cite{review}. One line of work extends traditional local-expansion frameworks to magnetic systems, as exemplified by magnetic moment tensor potentials (mMTP)\cite{mmtp} and magnetic variants of the atomic cluster expansion (ACE)\cite{magnetic_ace}, where magnetic moments are introduced as explicit variables in hand-crafted descriptors or systematically improvable many-body basis expansions. These approaches provide controllable and often interpretable representations, but their flexibility is closely connected to the chosen descriptor design and basis truncation. A second line of work introduces deep neural-network potentials for spin-lattice systems. For example, DeePSPIN\cite{deepspin} demonstrated that non-collinear magnetic potential energy surfaces can be learned in a unified deep-learning framework, notably through a pseudo-atom representation of magnetic degrees of freedom. More recently, equivariant message-passing models such as MAGNet\cite{magnet} and magnetic MACE (mMACE)\cite{mmace} have shown that continuous magnetic moments can be incorporated directly into symmetry-preserving neural architectures, substantially improving the modeling of non-collinear magnetism. SpinGNN++\cite{spingnn++} further highlighted the importance of embedding magnetic interaction structure and time-reversal-related constraints into the network design.

Despite this progress, data-efficient learning remains a central challenge for magnetic machine-learning potentials. Generating spin-polarized first-principles data spanning diverse magnetic states and atomic structures is computationally demanding, particularly when noncollinear spin configurations must be sampled jointly with variations in atomic positions and lattice parameters to describe spin-lattice coupling. Models trained on such data must also respect the required symmetries and resolve the subtle energy variations associated with magnetic interactions. Architectures without a representation tailored to the local coupling between a magnetic moment and its spin-lattice environment may require larger and more diverse datasets to learn the relevant magnetic interactions. This consideration motivates an approach that couples the central spin representation to an aggregated local spin-lattice environment and provides symmetry-allowed channels for describing various magnetic interactions.

In this work, we introduce the Spin Tensor Equivariant Potential (STEP), a magnetic machine-learning interatomic potential that treats local magnetic moments as continuous geometric degrees of freedom and embeds them in an equivariant representation space~\cite{e3nn}. Its key operation is a Center-Environment Tensor Product that couples the central spin representation to an aggregated local spin-lattice environment. The resulting coupled representation provides symmetry-allowed channels for learning isotropic exchange, antisymmetric exchange, and more general magnetic interactions. STEP preserves translational invariance and $\mathrm{SO}(3)$ equivariance, while global spin-reversal invariance is imposed at the energy level through feature symmetrization.

We evaluate STEP across empirical scaling tests, public benchmark datasets, and representative spin-lattice applications. Learning-curve analysis on monolayer CrI$_3$ shows that higher-order tensor channels and iterative center-environment couplings substantially improve data efficiency for energy, force, and magnetic force predictions. On public FeAl, CrN, and elemental Fe datasets~\cite{CrN_datasets,FeAl_datasets,deepspin_Fe}, STEP achieves competitive or improved accuracy compared with recent magnetic machine-learning potentials. We further demonstrate that STEP reproduces phonon and magnon spectra of monolayer CrI$_3$, captures magnon--phonon hybridization in Fe$_2$Mo$_3$O$_8$ semi-quantitatively, and drives large-scale spin-dynamics simulations yielding Curie temperatures for monolayer CrI$_3$ and body-centered-cubic (bcc) Fe close to experiment. These results establish STEP as a physically informed and data-efficient framework for modeling spin-lattice coupling, magnetic excitations, and finite-temperature magnetic behavior.

\section{\label{sec:methodology}The STEP Framework}

\begin{figure*}[htbp]
    \centering

    \includegraphics[width=0.95\textwidth]{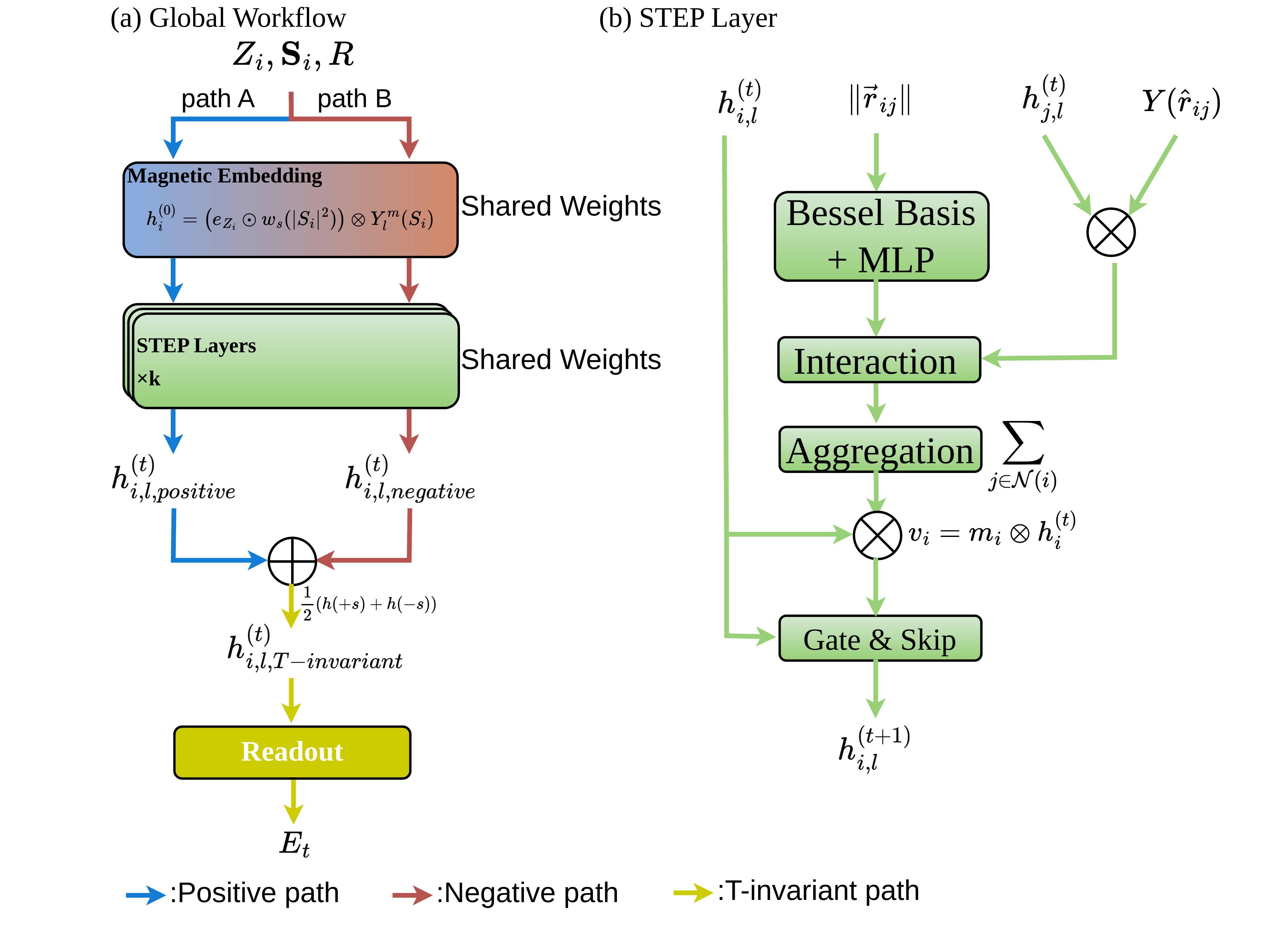}
    \caption{\label{fig:architecture} 
    \textbf{Overall architecture of the Spin Tensor Equivariant Potential (STEP).} 
    (a) The global workflow illustrates the concurrent processing of the positive ($+s$) and negative ($-s$) spin paths. The feature-level time-reversal symmetry is enforced by averaging the output representations before the layer-wise readouts. 
    (b) Detailed schematic of a single STEP layer, highlighting the Center-Environment Tensor Product (Center-Env TP) message passing. The model constructs physically meaningful many-body interactions by taking the geometric tensor product between the aggregated environmental field and the center atomic feature, naturally capturing complex spin-lattice couplings.
    }
\end{figure*}
\subsection{Magnetic Feature Embedding}

To integrate scalar chemical identities and continuous vector magnetic moments into a unified equivariant framework, STEP employs a magnetic embedding strategy guided by physical considerations. Unlike conventional graph neural networks that either separate scalar node features from directional vectors or rely on explicit normalization schemes, our approach maps the initial atomic states into a high-dimensional irreducible representation (irreps) space while remaining well-defined in the non-magnetic limit.

For an atom $i$ with atomic number $Z_i$ and local magnetic moment $\mathbf{S}_i$, the initial scalar chemical embedding $h_{\mathrm{chem}, i}^{(0)} \in \mathbb{R}^{C}$ is first generated via a learnable linear mapping of the one-hot encoded atomic species. In magnetic systems, especially itinerant magnets, the magnitude of the local magnetic moment is not necessarily rigidly fixed and may exhibit longitudinal spin fluctuations. To account for this behavior, we evaluate the squared magnitude of the spin vector, normalized by a system-specific reference scale $S_{\mathrm{ref}}$, and process it through a multilayer perceptron (MLP) to generate a nonlinear modulation weight:
\begin{equation}
w_s(|\mathbf{S}_i|^2) = \mathrm{MLP}\left(\frac{|\mathbf{S}_i|^2}{S_{\mathrm{ref}}^2}\right).
\end{equation}
By expressing this modulation as a function of $|\mathbf{S}_i|^2$ rather than $|\mathbf{S}_i|$, we avoid the non-differentiability associated with the norm at the origin. This choice is also consistent with the Landau-type perspective that thermodynamic quantities are often expanded in even powers of the magnetization near the non-magnetic limit \cite{landau}, thereby providing a smooth dependence of the embedding on the spin amplitude.

To encode the orientational information of the spin vector while preserving a smooth dependence on longitudinal spin fluctuations and the non-magnetic limit, STEP constructs the angular magnetic features directly from the scaled spin vector. Specifically, it computes the solid spherical harmonics $\mathcal{Y}_{\ell m}(\mathbf{S}_i)$ as:
\begin{equation}
\mathcal{Y}_{\ell m}(\tilde{\mathbf{S}}_i) = |\tilde{\mathbf{S}}_i|^\ell Y_{\ell m}\left(\frac{\tilde{\mathbf{S}}_i}{|\tilde{\mathbf{S}}_i|}\right), \quad \tilde{\mathbf{S}}_i = \frac{\mathbf{S}_i}{S_{\mathrm{ref}}}.
\end{equation}
In practice, the solid spherical harmonics are evaluated through their equivalent polynomial form in the Cartesian components of the spin vector, so that the representation remains well-defined at $\tilde{\mathbf{S}}_i = 0$\cite{e3nn,tfn}. Since solid spherical harmonics are homogeneous polynomials of degree $\ell$, they also provide a smooth geometric basis for subsequent tensor-product operations.

Finally, the scalar chemical feature modulated by the longitudinal fluctuation is aggregated with the solid spherical harmonics via a broadcasted tensor product to formulate the complete initial magnetic feature:
\begin{equation}
h_{i, (\ell,m,c)}^{(0)} = \left( h_{\mathrm{chem}, i,c}^{(0)} \odot w_{s,c}(|\mathbf{S}_i|^2) \right) \otimes \mathcal{Y}_{\ell m}(\tilde{\mathbf{S}}_i).
\label{eq:initial_feature}
\end{equation}
Here, $i$ indexes atoms, $c=1,\ldots,C$ labels feature channels, and $(\ell,m)$ labels the angular-momentum order and component of the spin spherical harmonics. The symbol $\odot$ denotes channel-wise multiplication, while the final product with $\mathcal{Y}_{\ell m}$ is implemented as a broadcasted tensor product over the channel and angular components.

This initialization couples the species-dependent chemical information with the symmetry properties of the magnetic degrees of freedom, providing the initial equivariant features for the subsequent high-order message passing.

\subsection{Center-Environment Tensor Product Message Passing}

While conventional $\mathrm{SO}(3)$-equivariant message-passing neural networks have demonstrated exceptional performance in modeling standard interatomic potentials\cite{nequip,mace}, they typically update the central atomic state by linearly superposing aggregated neighborhood messages. In magnetic systems, however, the potential energy surface is shaped by nontrivial couplings between local magnetic degrees of freedom and their spin-lattice environment, which are not naturally captured by simple additive updates alone. To systematically capture this underlying physics, our interaction block is constructed to perform Center-Environment Tensor Product message passing while maintaining strict $\mathrm{SO}(3)$ equivariance.

At layer $t$, the feature associated with atom $i$ is denoted by
\begin{equation}
h_i^{(t)} \in \mathcal{F}
:= \bigoplus_{\ell=0}^{\ell_{\max}} \mathbb{R}^{m_\ell} \otimes V_\ell,
\label{eq:feature_space}
\end{equation}
where $V_\ell$ is the irreducible representation of $\mathrm{SO}(3)$ of angular momentum order $\ell$, and $m_\ell$ is its multiplicity. The subsequent interaction layers preserve this rotationally organized feature structure up to learnable equivariant linear maps. The direct sum $\oplus$ indicates that features of different angular orders are stored as separate irreducible representation channels.

For an edge $j \to i$, where $i$ denotes the central atom and $j\in\mathcal{N}(i)$ denotes one of its neighboring atoms, we define the relative spatial geometry
\begin{equation}
\begin{split}
\mathbf{r}_{ij} &= \mathbf{x}_j - \mathbf{x}_i + \mathbf{s}_{ij}, \\
r_{ij} &= \|\mathbf{r}_{ij}\|, \\
\hat{\mathbf{r}}_{ij} &= \frac{\mathbf{r}_{ij}}{r_{ij}},
\end{split}
\label{eq:edge_geometry}
\end{equation}
where $\mathbf{s}_{ij}$ is the periodic-image shift. The geometric edge message $m_{ij}^{(t)}$ transmitted from neighbor $j$ to central atom $i$ is abstractly computed as
\begin{equation}
m_{ij}^{(t)}
=
\mathcal{T}^{(t)}
\!\left(
W^{(t)} h_j^{(t)},
\,Y(\hat{\mathbf{r}}_{ij}),
\,\omega^{(t)}(r_{ij})
\right),
\label{eq:edge_message_so3}
\end{equation}
where $W^{(t)}$ is an $\mathrm{SO}(3)$-equivariant linear map, $Y(\hat{\mathbf{r}}_{ij})$ denotes the spherical-harmonic basis on the edge direction, and $\omega^{(t)}(r_{ij})$ is a learnable radial weight. In practice, this mechanism evaluates the algebraic tensor multiplication explicitly using the Clebsch-Gordan coefficients $C$ \cite{tfn,e3nn}:
\begin{equation}
\begin{split}
m_{ij,(\ell_o, m_o, c)}^{(t)} &= \omega^{(t)}(r_{ij})_{c} \\
&\quad \times \sum_{m_i, m_f} C_{(\ell_i, m_i), (\ell_f, m_f)}^{(\ell_o, m_o)} \tilde{h}_{j, (\ell_i, m_i, c)}^{(t)} Y_{\ell_f}^{m_f}(\hat{\mathbf{r}}_{ij}).
\end{split}
\label{eq:cg_tensor_product}
\end{equation}
Here, $(\ell_i,m_i)$, $(\ell_f,m_f)$, and $(\ell_o,m_o)$ label the input feature, edge spherical-harmonic, and output irrep components, respectively, while $c$ denotes the corresponding feature channel.
In this expression, $\tilde{h}_{j}^{(t)} = W^{(t)} h_j^{(t)}$, and the radial weight $\omega^{(t)}(r_{ij})_{c}$ acts as the physical analogue of the distance-dependent exchange couplings (e.g., $J(r_{ij})$). It is dynamically parameterized by a continuous MLP operating on a Bessel basis expansion of the interatomic distance. Subsequently, these pairwise messages are explicitly normalized by the square root of the average neighbor count $\sqrt{\langle N \rangle}$ to maintain numerical stability during deep network propagation, yielding the aggregated local environment field:
\begin{equation}
m_i^{(t)} = \sum_{j \in \mathcal{N}(i)} \frac{1}{\sqrt{\langle N \rangle}} m_{ij}^{(t)}.
\label{eq:environment_feature}
\end{equation}
Here, $\mathcal{N}(i)$ denotes the neighbor list of atom $i$, and $\langle N\rangle$ is the average number of neighbors used for normalization.

The central operation of the layer is a bilinear coupling between the aggregated environment feature and the central on-site feature,
\begin{equation}
v_i^{(t)}
=
\mathcal{B}^{(t)}
\!\left(
L_{\mathrm{env}}^{(t)} m_i^{(t)},
\,L_{\mathrm{ctr}}^{(t)} h_i^{(t)}
\right),
\label{eq:center_environment_bilinear}
\end{equation}
where $L_{\mathrm{env}}^{(t)}$ and $L_{\mathrm{ctr}}^{(t)}$ are $\mathrm{SO}(3)$-equivariant linear maps and $\mathcal{B}^{(t)}$ is a learnable equivariant bilinear map realized by a fully connected tensor product. Equation~\eqref{eq:center_environment_bilinear} should be interpreted as a coupling between \emph{environmental covariant features} and \emph{central covariant features}. This operation provides a symmetry-constrained bilinear basis in the latent feature space, from which the network can learn the complex combinations relevant to the target magnetic energy landscape.

The representation-theoretic motivation of Eq.~\eqref{eq:center_environment_bilinear} is straightforward. Whenever $\ell=1$ channels are present in the coupled features, the tensor-product decomposition
\begin{equation}
V_1 \otimes V_1 \cong V_0 \oplus V_1 \oplus V_2
\label{eq:vector_tensor_decomp}
\end{equation}
implies that the bilinear interaction naturally generates scalar, vector, and rank-two channels. In Cartesian notation, for any two vectors $\mathbf{a},\mathbf{b}\in\mathbb{R}^3$,
\begin{equation}
\begin{split}
\mathbf{a}\otimes\mathbf{b}
&= \underbrace{\frac{1}{3}(\mathbf{a}\cdot\mathbf{b}) I}_{\ell=0} 
+ \underbrace{\frac{1}{2}\left(\mathbf{a}\otimes\mathbf{b}-\mathbf{b}\otimes\mathbf{a}\right)}_{\ell=1} \\
&\quad + \underbrace{
\frac{1}{2}\left(\mathbf{a}\otimes\mathbf{b}+\mathbf{b}\otimes\mathbf{a}\right)
-\frac{1}{3}(\mathbf{a}\cdot\mathbf{b})I
}_{\ell=2}.
\end{split}
\label{eq:cartesian_irreducible_decomp}
\end{equation}
Thus, within the latent space, the center-environment tensor coupling naturally provides the irreducible rotational channels needed to encode isotropic scalar couplings, antisymmetric $\ell=1$ couplings, and symmetric rank-two anisotropic couplings. While the decomposition of vector channels ($V_1 \otimes V_1$) provides an intuitive parallel to conventional two-spin interactions, it is important to emphasize that our implementation fully evaluates tensor products across all supported representation orders (e.g., $V_2 \otimes V_1$, $V_2 \otimes V_2$). These higher-order channels systematically expand the latent space beyond simple pairwise vector couplings, providing the representational capacity required to capture complex local anisotropies and higher-order many-body spin interactions.

\subsection{Equivariant Gating and Residual Interaction Update}

To introduce additional nonlinear expressivity beyond equivariant linear maps and bilinear tensor products, the bilinearly coupled feature $v_i^{(t)}$ is passed through an equivariant gated nonlinearity \cite{gating_cnn}. Let
\begin{equation}
v_i^{(t)} = v_{i,\mathrm{sc}}^{(t)} \oplus v_{i,\mathrm{ns}}^{(t)},
\end{equation}
where $v_{i,\mathrm{sc}}^{(t)}$ collects all true scalar channels ($\ell=0$) and $v_{i,\mathrm{ns}}^{(t)}$ collects all non-scalar tensor channels ($\ell>0$). The gated nonlinearity $\Phi^{(t)}$ has the generic form:
\begin{equation}
\Phi^{(t)}\!\left(v_i^{(t)}\right)
=
\phi_{\mathrm{sc}}^{(t)}\!\left(v_{i,\mathrm{sc}}^{(t)}\right)
\oplus
\left[
\phi_{\mathrm{g}}^{(t)}\!\left(g_i^{(t)}\right)
\odot
v_{i,\mathrm{ns}}^{(t)}
\right],
\end{equation}
where $\odot$ denotes channel-wise multiplication. Specifically, for the pure scalar channels ($\ell=0$), the network applies standard smooth nonlinearities, implemented here using the SiLU activation function $\phi_{\mathrm{sc}}^{(t)}$. For non-scalar tensor channels ($\ell>0$), applying standard component-wise nonlinear functions directly would in general break rotational equivariance. To avoid this issue, the preceding linear layer outputs an auxiliary set of scalar gates $g_i^{(t)}$, which are mapped to the $(0,1)$ range via a sigmoid activation, i.e., $\phi_{\mathrm{g}}^{(t)}(g_i^{(t)}) = \sigma(g_i^{(t)})$. Since multiplication by an $\mathrm{SO}(3)$-invariant scalar commutes with the group action on each non-scalar irrep channel, this gating mechanism introduces nonlinear representational capacity while strictly preserving continuous rotational equivariance.

Finally, the nonlinearly activated interaction features are integrated with the pre-interaction state through a residual skip connection to formulate the layer update:
\begin{equation}
h_i^{(t+1)}
=
P^{(t)} \Phi^{(t)}\!\left(v_i^{(t)}\right)
+
S^{(t)} h_i^{(t)},
\label{eq:layer_update_so3}
\end{equation}
with $P^{(t)}$ and $S^{(t)}$ denoting learnable $\mathrm{SO}(3)$-equivariant linear maps. The second term provides a residual update that propagates transformed central-atom features in parallel with the center-environment contribution. This skip pathway preserves low-order local information, including scalar magnetic-amplitude features, and improves optimization stability by allowing such information to be carried directly across layers.

\subsection{Feature-Level Time-Reversal Symmetry and Energy Readout}

In the absence of external time-reversal-odd fields, the magnetic potential energy should be invariant under a global reversal of all magnetic moments,
\begin{equation}
E(\{\mathbf{S}_i\}) = E(\{-\mathbf{S}_i\}) .
\label{eq:trs_energy_condition}
\end{equation}
To impose this constraint at the model level, STEP employs a feature-level time-reversal symmetrization before the final energy readout.

Let $h_i^{(t)}(\{\mathbf{S}\})$ denote the hidden feature of atom $i$ at layer $t$ obtained from a forward pass with magnetic moments $\{\mathbf{S}\}$. Since the total energy is a rotational scalar, the energy readout is performed only on the $\ell=0$ channels. We denote the projection onto these scalar channels by $\Pi_0$. To enforce global time-reversal symmetry, STEP evaluates the scalar features for both the original magnetic configuration and its globally reversed counterpart, and then averages them:
\begin{equation}
\bar{z}_i^{(t)}
=
\frac{1}{2}
\left[
\Pi_0 h_i^{(t)}(\{\mathbf{S}\})
+
\Pi_0 h_i^{(t)}(\{-\mathbf{S}\})
\right].
\label{eq:trs_scalar_symmetrization}
\end{equation}
This operation removes the time-reversal-odd component of the scalar latent representation before it enters the energy readout. Equivalently, the symmetrized feature satisfies
\begin{equation}
\bar{z}_i^{(t)}(\{\mathbf{S}\})
=
\bar{z}_i^{(t)}(\{-\mathbf{S}\}) ,
\label{eq:trs_feature_even}
\end{equation}
which is sufficient to make any subsequent scalar readout invariant under global reversal of the magnetic moments.

The atomic energy is obtained using a layer-wise scalar readout. For each interaction depth $t$, an independent multilayer perceptron $f^{(t)}$ maps the symmetrized scalar feature $\bar{z}_i^{(t)}$ to an atomic energy contribution:
\begin{equation}
E_i
=
\sum_{t=0}^{L}
f^{(t)}
\left(
\bar{z}_i^{(t)}
\right).
\label{eq:atomic_energy_sum}
\end{equation}
The total energy is then written as
\begin{equation}
E
=
\sum_i E_i
+
\sum_i E_{z_i}^{\mathrm{shift}},
\label{eq:total_energy}
\end{equation}
where $E_{z_i}^{\mathrm{shift}}$ is a species-dependent reference energy shift.

Because the same readout functions are applied to the symmetrized scalar features and because Eq.~\eqref{eq:trs_feature_even} holds by construction, the predicted energy satisfies
\begin{equation}
E(\{\mathbf{S}_i\}) = E(\{-\mathbf{S}_i\})
\end{equation}
exactly, up to numerical precision. This symmetry is imposed at the feature level rather than through data augmentation.

The layer-wise readout provides a physically motivated hierarchy of magnetic energy contributions. The $t=0$ term accesses the initial local magnetic embedding, including scalar information associated with the on-site magnetic amplitude before any neighborhood convolution. Deeper readout terms incorporate progressively enriched spin-lattice environments generated by the equivariant message-passing and center-environment tensor-product interactions. In this way, STEP combines local Landau-like magnetic-amplitude contributions with environment-dependent exchange-like and many-body spin-lattice couplings within a unified symmetry-preserving energy model.

\subsection{\label{sec:training}Training and Optimization}

STEP is trained using a joint loss function that combines the total energy, forces, unit-cell virial tensor, and magnetic forces. Specifically, the total loss $\mathcal{L}$ is defined as a weighted sum of the mean absolute error (MAE) terms between the density functional theory (DFT) reference values and the corresponding model predictions:
\begin{equation}
    \mathcal{L}
    =
    \lambda_E \mathcal{L}_E
    +
    \lambda_F \mathcal{L}_F
    +
    \lambda_V \mathcal{L}_V
    +
    \lambda_M \mathcal{L}_M,
\end{equation}
where $\lambda_E$, $\lambda_F$, $\lambda_V$, and $\lambda_M$ are system-specific weights used to balance the contributions of the different target quantities. To ensure that the predicted derivatives are consistent with the learned potential energy surface, the forces $\hat{\mathbf{F}}_i$, virial tensor $\hat{\mathcal{W}}$, and magnetic forces $\hat{\mathbf{F}}_j^M$ are obtained from the predicted total energy $\hat{E}$ by automatic differentiation~\cite{nequip,mace}:
\begin{equation}
    \hat{\mathbf{F}}_i
    =
    -\frac{\partial \hat{E}}{\partial \mathbf{r}_i},
    \qquad
    \hat{\mathcal{W}}
    =
    -\frac{\partial \hat{E}}{\partial \boldsymbol{\epsilon}},
    \qquad
    \hat{\mathbf{F}}_j^M
    =
    -\frac{\partial \hat{E}}{\partial \mathbf{M}_j},
\end{equation}
where $\mathbf{r}_i$, $\boldsymbol{\epsilon}$, and $\mathbf{M}_j$ denote the atomic position, global strain tensor, and local magnetic moment, respectively. Throughout this work, ``force'' refers to the negative energy gradient with respect to the atomic position, whereas ``magnetic force'' refers to the negative energy gradient with respect to the local magnetic moment. The individual loss components are defined as follows:
\begin{align}
    \mathcal{L}_E &= \frac{1}{N} \left| \hat{E} - E \right|, \\
    \mathcal{L}_F &= \frac{1}{3N} \sum_{i=1}^{N} \sum_{\alpha \in \{x,y,z\}} \left| \hat{F}_{i,\alpha} - F_{i,\alpha} \right|, \\
    \mathcal{L}_V &= \frac{1}{9} \sum_{\alpha, \beta \in \{x,y,z\}} \left| \hat{\mathcal{W}}_{\alpha\beta} - \mathcal{W}_{\alpha\beta} \right|, \\
    \mathcal{L}_M &= \frac{1}{3N_{mag}} \sum_{j=1}^{N_{mag}} \sum_{\alpha \in \{x,y,z\}} \left| \hat{F}^{M(\perp)}_{j,\alpha} - F^{M(\perp)}_{j,\alpha} \right|,
\end{align}
where $N$ and $N_{mag}$ denote the total number of atoms and the number of magnetic atoms in the structure, respectively, while $E$, $F_{i,\alpha}$, $\mathcal{W}_{\alpha\beta}$, and $F^M_{j,\alpha}$ are the corresponding DFT reference values, the parenthesized superscript $(\perp)$ indicates an
optional projection.

The magnetic force in STEP is defined as the negative energy gradient with respect to the full local magnetic moment vector, so it naturally contains both longitudinal and transverse components. Accordingly, the magnetic-force loss $\mathcal{L}_M$ can be evaluated using the full vector magnetic force in the general case. For rigid-spin applications, where the local moment magnitude is constrained and only transverse spin fluctuations are active, we can optionally use the projected magnetic force
\[
\mathbf{F}^{M\perp}
=
\mathbf{F}^{M}
-
(\mathbf{F}^{M}\cdot \mathbf{e}_m)\mathbf{e}_m,
\]
where $\mathbf{e}_m=\mathbf{M}/|\mathbf{M}|$ is the unit vector along the local magnetic moment. This projected loss penalizes the transverse components relevant to Landau-Lifshitz spin dynamics\cite{llg}, while the full-vector formulation remains available for systems with longitudinal moment fluctuations.

The model parameters are optimized using the Adam optimizer\cite{adam} with the AMSGrad\cite{amsgrad} variant. 
To improve training stability, we employ a plateau-based learning-rate scheduler, which reduces the learning rate by a predefined decay factor when the validation loss shows no improvement for a specified number of consecutive epochs (patience). 
An early-stopping criterion is also used to terminate training once the validation loss no longer improves, thereby mitigating overfitting.
The specific hyperparameters---including the initial learning rate, batch size, loss weights ($\lambda_E, \lambda_F, \lambda_V, \lambda_M$), and scheduler patience---are tuned separately for different target materials. 
Such system-dependent configuration is motivated not only by differences in dataset size and convergence behavior, but also by differences in the dominant underlying physical mechanisms. 
For example, the training setup for two-dimensional CrI$_3$ should account for the stronger role of spin-orbit coupling and local magnetic anisotropy, whereas that for three-dimensional bulk Fe must better accommodate its more itinerant magnetic character.
This flexibility in the joint loss formulation also enables property-specific optimization. 
For instance, the study of local harmonic vibrational properties (e.g., phonon spectra) benefits from high accuracy in restoring forces and virials near equilibrium, whereas simulations of finite-temperature magnetic phase transitions (e.g., Curie temperature) place greater emphasis on the accuracy of the spin-dependent energy landscape. 
By adjusting the relative loss weights accordingly, the training can be biased toward the physical observables most relevant to the target application.
Finally, global random seeds are fixed to ensure reproducible dataset splitting and network initialization.

To further augment the numerical stability and the out-of-distribution generalization capability of the model, several advanced stabilization techniques are integrated. We incorporate an exponential moving average (EMA) of the model weights\cite{ema}. During the validation phase and final inference, these smoothed EMA weights are evaluated, effectively mitigating the optimization noise inherent in complex highly non-linear energy landscapes. Gradient clipping is also systematically applied to prevent exploding gradients during backpropagation. Prior to training, the network undergoes a physics-informed, data-driven initialization: the baseline energy shift and the root-mean-square (RMS) force scale are pre-computed directly from the training set. Additionally, to stabilize the computation of solid spherical harmonics, the input magnetic moments are dynamically scaled by a material-specific reference value $S_{\text{ref}}$. This reference is derived from the maximum atomic magnetic moment present in the dataset augmented by a safety padding factor, ensuring that the high-order polynomial features remain numerically well-behaved throughout the training lifecycle.

\section{Results}

\subsection{Benchmarks on public datasets}

\subsubsection{FeAl}

\begin{figure*}[htbp]
    \centering
    \includegraphics[width=0.99\linewidth]{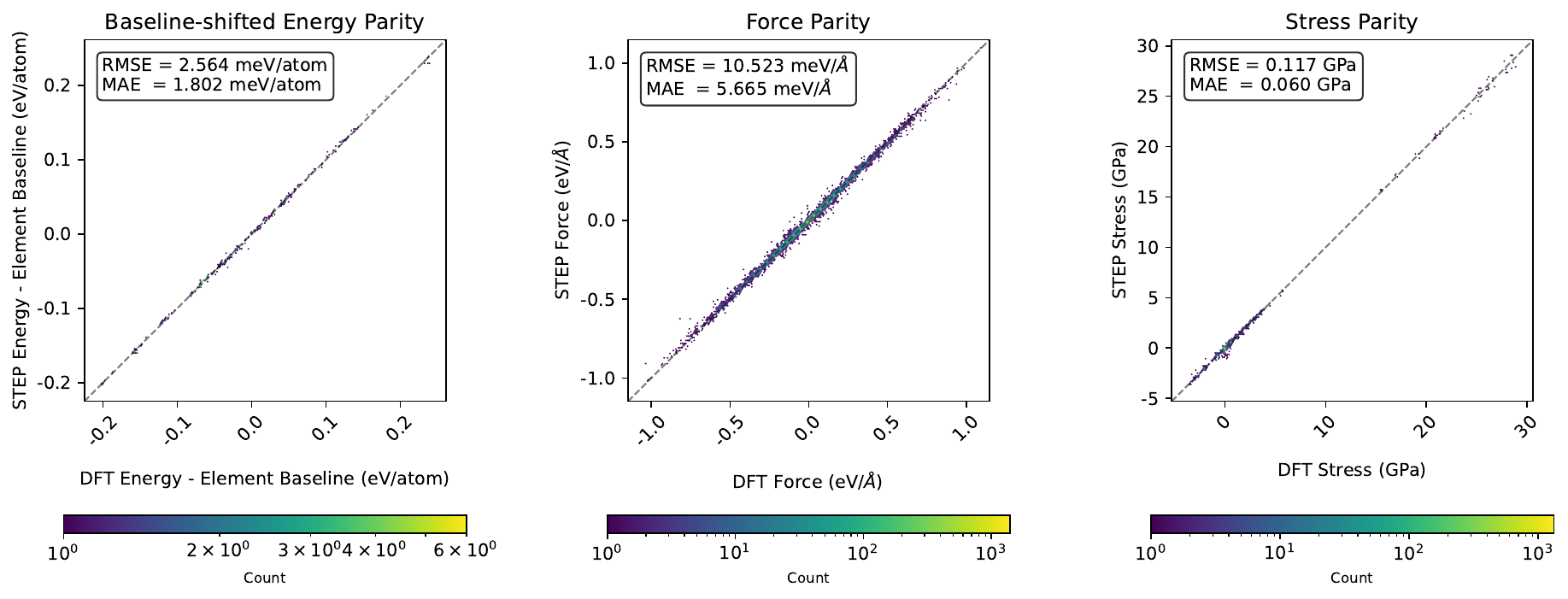}
    \caption{
    Parity plots of STEP on the FeAl benchmark dataset.
    The predicted energy, forces, and stress are compared with the corresponding DFT reference values.
    }
    \label{fig:feal_parity}
\end{figure*}

\begin{figure*}[htbp]
    \centering
    \includegraphics[width=0.99\linewidth]{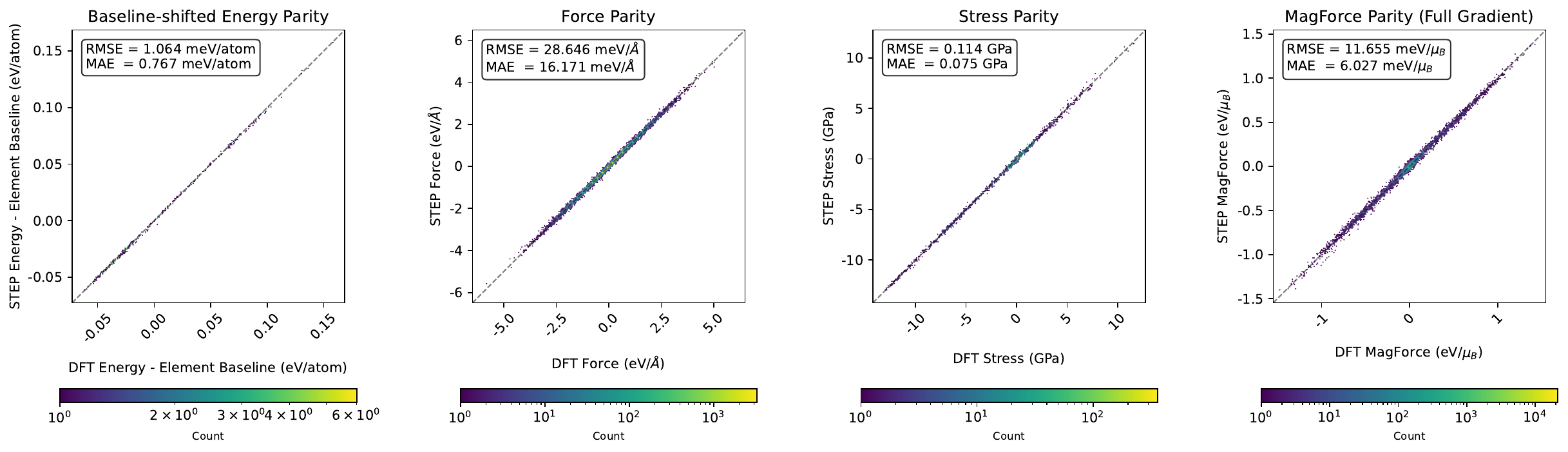}
    \caption{
    Parity plots of STEP on the CrN benchmark dataset.
    The predicted energy, forces, stress, and magnetic forces are compared with the corresponding DFT reference values.
    }
    \label{fig:crn_parity}
\end{figure*}

\begin{figure*}[htbp]
    \centering
    \includegraphics[width=0.99\linewidth]{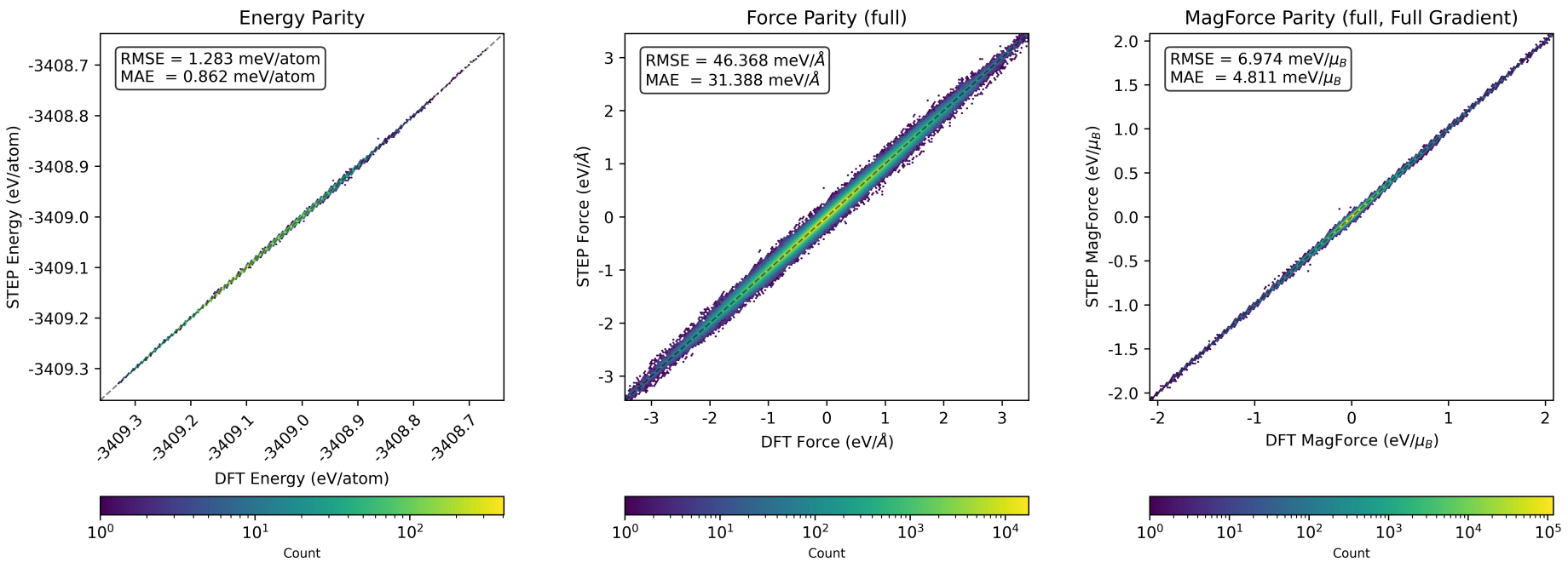}
    \caption{
    Parity plots of STEP-full over the complete original Fe dataset. The predicted energy, forces, and magnetic forces are compared with the corresponding DFT reference values.
    }
    \label{fig:bcc_fe_parity}
\end{figure*}

We evaluate STEP on the publicly available FeAl benchmark dataset \cite{FeAl_datasets} and compare its performance with the reported results of mMTP\cite{mmtp}, MACE\cite{mace}, and mMACE\cite{mmace}. To facilitate comparison, we choose the architectural hyperparameters of STEP to be as close as possible to those used in the reported benchmark, especially in terms of hidden feature dimension, angular resolution, radial resolution, and interaction cutoff.

Specifically, STEP uses three message-passing layers with 64 hidden channels. The spatial spherical harmonics are expanded up to $l_{\mathrm{pos}}^{\max}=3$, and the magnetic-moment spherical harmonics are expanded up to $l_{\mathrm{mag}}^{\max}=1$. The radial embedding uses 8 Bessel basis functions with a polynomial cutoff, and the cutoff radius is set to $r_{\mathrm{cut}}=5.0~\mathrm{\AA}$. The radial network and the layer-wise energy readout network both use a hidden width of 64. Only Fe atoms are treated as magnetic atoms, while Al atoms are treated as non-magnetic atoms.

Figure~\ref{fig:feal_parity} shows the parity plots of the predicted energy, force, and stress against the DFT reference values on the FeAl dataset. The predictions lie close to the diagonal, indicating that STEP accurately reproduces the DFT energy landscape and its derivatives.
Table~\ref{tab:feal_benchmark_comparison} compares the root-mean-square errors (RMSEs) obtained by STEP with the corresponding RMSEs reported for mMTP \cite{mmtp}, MACE\cite{mace}, and mMACE\cite{mmace} on the FeAl benchmark. STEP achieves the lowest errors in energy, force, and stress among the compared methods.

\begin{table}[H]
\centering
\caption{Benchmark comparison on the FeAl dataset. }
\begin{tabular*}{\linewidth}{@{\extracolsep{\fill}}lcccc}
\toprule
RMSE & mMTP & MACE & mMACE & STEP \\
\midrule
Energy (meV/atom)              & 4.03 & 13.86 & 2.69 & \textbf{2.564} \\
Force (meV/$\mathrm{\AA}$)    & 48   & 52    & 11   & \textbf{10.523} \\
Stress (GPa)                   & 0.79 & 0.93  & 0.26 & \textbf{0.117} \\
Magnetic force (meV/$\mu_B$)   & --   & --   & --   & -- \\
\bottomrule
\end{tabular*}
\label{tab:feal_benchmark_comparison}
\end{table}

\subsubsection{CrN}

We further evaluate STEP on the publicly available CrN benchmark dataset\cite{CrN_datasets} and compare its performance with the reported results of mMTP\cite{mmtp}, MACE\cite{mace}, and mMACE\cite{mmace}. To maintain consistency and facilitate comparison, we use the identical capacity-controlling architectural hyperparameters as detailed for the benchmark on FeAl, including the interaction cutoff, hidden feature dimension, and angular resolution. Cr atoms are treated as magnetic atoms, while N atoms are treated as non-magnetic atoms.

Figure~\ref{fig:crn_parity} shows the parity plots of the predicted energy, force, stress, and magnetic force against the DFT reference values on the CrN dataset. The predictions are close to the diagonal, indicating that STEP can accurately reproduce the DFT energy landscape, including forces, stresses, and magnetic forces. Table~\ref{tab:crn_benchmark_comparison} compares the RMSEs obtained by STEP with the corresponding values reported for mMTP\cite{mmtp}, MACE\cite{mace}, and mMACE\cite{mmace} on the CrN benchmark. STEP achieves the lowest errors in energy, stress, and magnetic force, while maintaining a force error close to mMACE.

\begin{table}[H]
\centering
\caption{Benchmark comparison on the CrN dataset. }
\begin{tabular}{lcccc}
\toprule
RMSE & mMTP & MACE & mMACE & STEP \\
\midrule
Energy (meV/atom)              & 1.7  & 31.2 & 1.21 & \textbf{1.064} \\
Force (meV/$\mathrm{\AA}$)    & 108  & 219  & \textbf{24} & 28.646 \\
Stress (GPa)                   & 0.4  & 1.29 & 0.25 & \textbf{0.114} \\
Magnetic force (meV/$\mu_B$)   & 64   & --   & 19   & \textbf{11.655} \\
\bottomrule
\end{tabular}
\label{tab:crn_benchmark_comparison}
\end{table}

\subsubsection{Fe}

To evaluate STEP on elemental Fe, a prototypical itinerant magnetic system, we use the publicly available dataset reported in Ref.~\cite{deepspin_Fe}. The dataset contains approximately 35,000 configurations generated through structural and magnetic perturbations and active learning, covering diverse atomic environments and variations in both orientations and magnitudes of local Fe magnetic moments. We train two models with the same architecture: STEP-full uses the complete dataset with a 90:10 training--validation split, whereas STEP-5k uses 5,000 representative configurations selected from the complete dataset by principal component analysis (PCA)\cite{pca} followed by farthest-point sampling (FPS)\cite{fps}, divided into 4,500 training and 500 validation configurations. In both cases, the validation configurations are excluded from gradient-based parameter updates.

Both models employ two message-passing layers, 8 hidden channels, magnetic and latent equivariant features up to $l_{\max}=2$, and a cutoff radius of $4.5~\mathrm{\AA}$. They are evaluated over the complete original dataset using identical RMSE definitions. The parity plots correspond to STEP-full, while all subsequent phonon, magnon, and finite-temperature calculations use the same STEP-5k checkpoint without task-specific retraining.

Figure~\ref{fig:bcc_fe_parity} demonstrates close agreement between the STEP-full predictions and the DFT reference values for energy, force, and magnetic force across the complete original dataset. Table~\ref{tab:bcc_fe_benchmark_comparison} compares the RMSEs of STEP-full and STEP-5k evaluated over the complete original dataset with the training errors reported for DeePSPIN-DZP~\cite{deepspin_Fe}. STEP-full achieves the lowest errors. Reducing the model-development subset to 5,000 representative configurations increases the corresponding STEP-5k errors. Nevertheless, these STEP-5k errors remain lower than the reported DeePSPIN-DZP training errors by approximately 70.9\%, 42.4\%, and 70.5\% for energy, force, and magnetic force, respectively.

Because the results were not obtained using identical data partitions and evaluation protocols, this comparison is not a strictly controlled head-to-head benchmark. Nevertheless, the low RMSEs of STEP-5k when evaluated over the complete original dataset provide complementary evidence of STEP's practical sample efficiency.

\begin{table}[htbp]
\centering
\caption{Benchmark comparison on the complete original Fe dataset. The DeePSPIN-DZP values are the reported training errors, whereas both STEP models are evaluated over the complete original dataset.}
\label{tab:bcc_fe_benchmark_comparison}

\small
\begin{tabular*}{\columnwidth}{@{\extracolsep{\fill}}lccc}
\toprule
RMSE
& \shortstack{DeePSPIN-\\DZP}
& STEP-full
& STEP-5k \\
\midrule
$E$ (meV/atom)
& 6.88
& \textbf{1.283}
& 1.999 \\
$F$ (meV/$\mathrm{\AA}$)
& 106.0
& \textbf{46.368}
& 61.014 \\
$F^{M}$ (meV/$\mu_B$)
& 28.6
& \textbf{6.974}
& 8.440 \\
\bottomrule
\end{tabular*}
\end{table}

\begin{figure}[htbp]
\centering
\includegraphics[width=0.88\columnwidth]{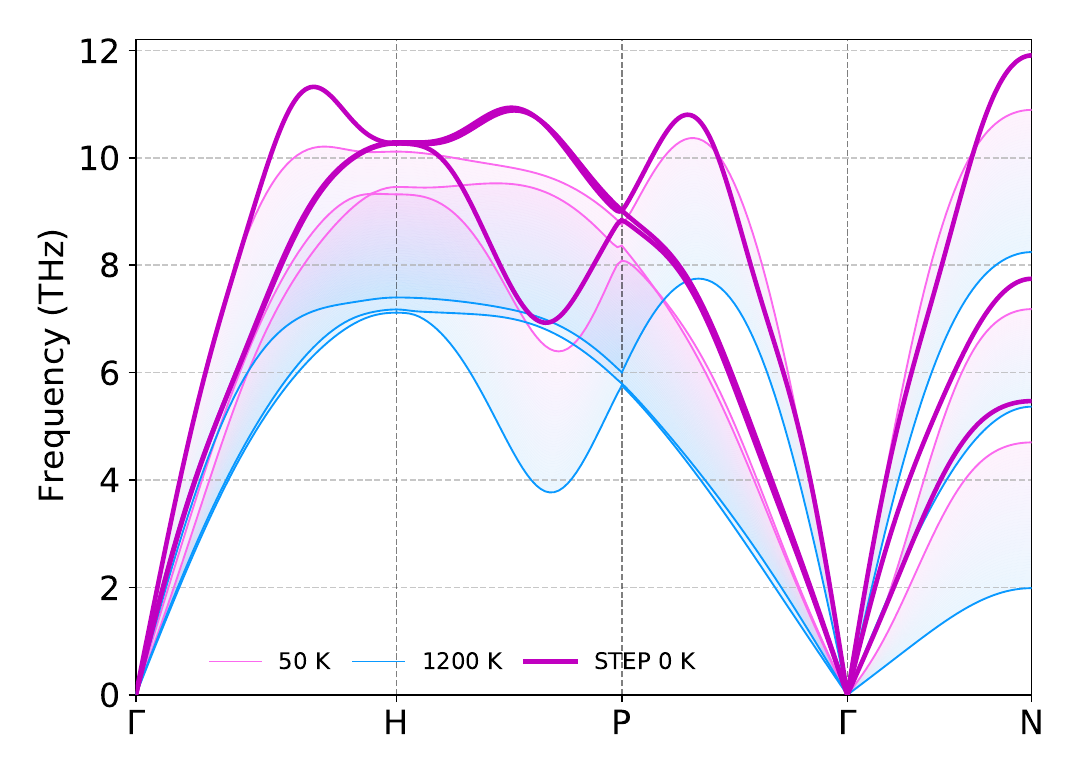}
\caption{
Harmonic phonon dispersion of bcc Fe predicted by STEP at 0 K,
compared with finite-temperature reference spectra at 50 K and
1200 K from an unpublished study~\cite{Xu2026Private}.
}
\label{fig:bcc_fe_phonon}
\end{figure}

\begin{figure}[htbp]
    \centering
    \includegraphics[width=0.86\columnwidth]{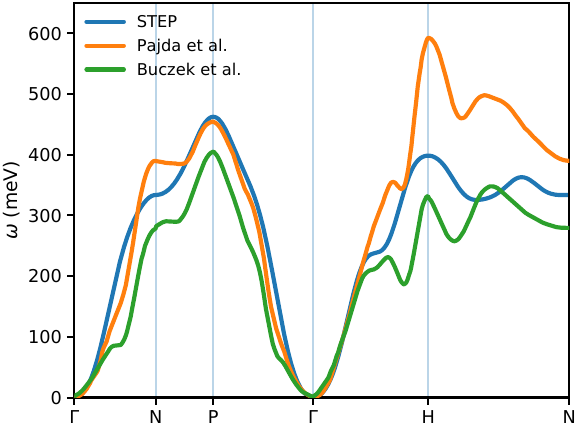}
    \caption{Magnon dispersion of ferromagnetic bcc Fe along the common high-symmetry path $\Gamma$--N--P--$\Gamma$--H--N. The STEP prediction is compared with the magnetic-force-theorem results reported by Pajda \emph{et al.}~\cite{pajda2001} and Buczek \emph{et al.}~\cite{buczek2011}. For direct comparison, the reference curves are replotted along the same sequence of symmetry-line segments, with the point denoted by $\Pi$ in Ref.~\cite{buczek2011} labeled as P here.}
    \label{fig:bcc_fe_magnon}
\end{figure}

We further examine collective lattice excitations in bcc Fe. As shown in Fig.~\ref{fig:bcc_fe_phonon}, the harmonic phonon dispersion predicted by STEP at 0 K captures the overall shape of the phonon branches along $\Gamma$--H--P--$\Gamma$--N. The STEP spectrum is systematically stiffer than the finite-temperature reference spectra at 50 K and 1200 K~\cite{Xu2026Private}, consistent with the expected phonon softening at elevated temperatures. Because the reference spectra include finite-temperature anharmonic and magnetic-disorder effects, this comparison is intended to be qualitative. Nevertheless, the physically reasonable phonon dispersion indicates that STEP provides a reliable description of the local curvature of the energy surface with respect to atomic displacements in bcc Fe.

We further examine the magnon dispersion of ferromagnetic bcc Fe predicted by STEP, as shown in Fig.~\ref{fig:bcc_fe_magnon}. For comparison, we include the results by Pajda \emph{et al.}~\cite{pajda2001} and Buczek \emph{et al.}~\cite{buczek2011}, which were obtained based on the Heisenberg model and magnetic force theorem (MFT)\cite{MFT}. Noticeable differences are present not only between STEP and the MFT results, but also between the two MFT results themselves, particularly in the high-energy region and around the H point.

Relative to the result of Buczek \emph{et al.}, STEP generally predicts higher magnon energies over much of the Brillouin-zone path, whereas it predicts a softer dispersion near H compared with the result of Pajda \emph{et al.} This discrepancy may arise from the fact that the magnetic force theorem is based on the Heisenberg model, whereas our approach does not rely on any empirical model Hamiltonian. In addition, Pajda \emph{et al.}~\cite{pajda2001} attributed the local minimum near H to a Kohn anomaly induced by long-range Ruderman-Kittel-Kasuya-Yoshida (RKKY) interactions and showed that this region is sensitive to the number of exchange shells included, with the calculated magnon bandwidth at H changing from 441 meV to 550 meV when the number of shells was increased from 15 to 172. The differences among STEP and the two MFT results reflect the sensitivity of the high-energy magnon spectra to details of methodology. Nevertheless, the agreement in the overall energy scale and dispersion structure indicates that STEP captures the magnetic interactions governing spin-wave excitations in bcc Fe.

Taken together, the STEP-full parity plots and the comparison among DeePSPIN-DZP, STEP-full, and STEP-5k establish the accuracy of STEP on the public Fe dataset. More importantly, the compact-data model retains low RMSEs evaluated over the complete original dataset and captures the main features of both phonon and magnon dispersions despite being developed from only 5,000 representative configurations. Together with the finite-temperature results below, these findings provide complementary evidence of the sample efficiency and physical transferability of STEP.

\subsection{Data Efficiency and Empirical Scaling Analysis}

\begin{figure*}[htbp]
    \centering
    \includegraphics[width=0.8\linewidth]{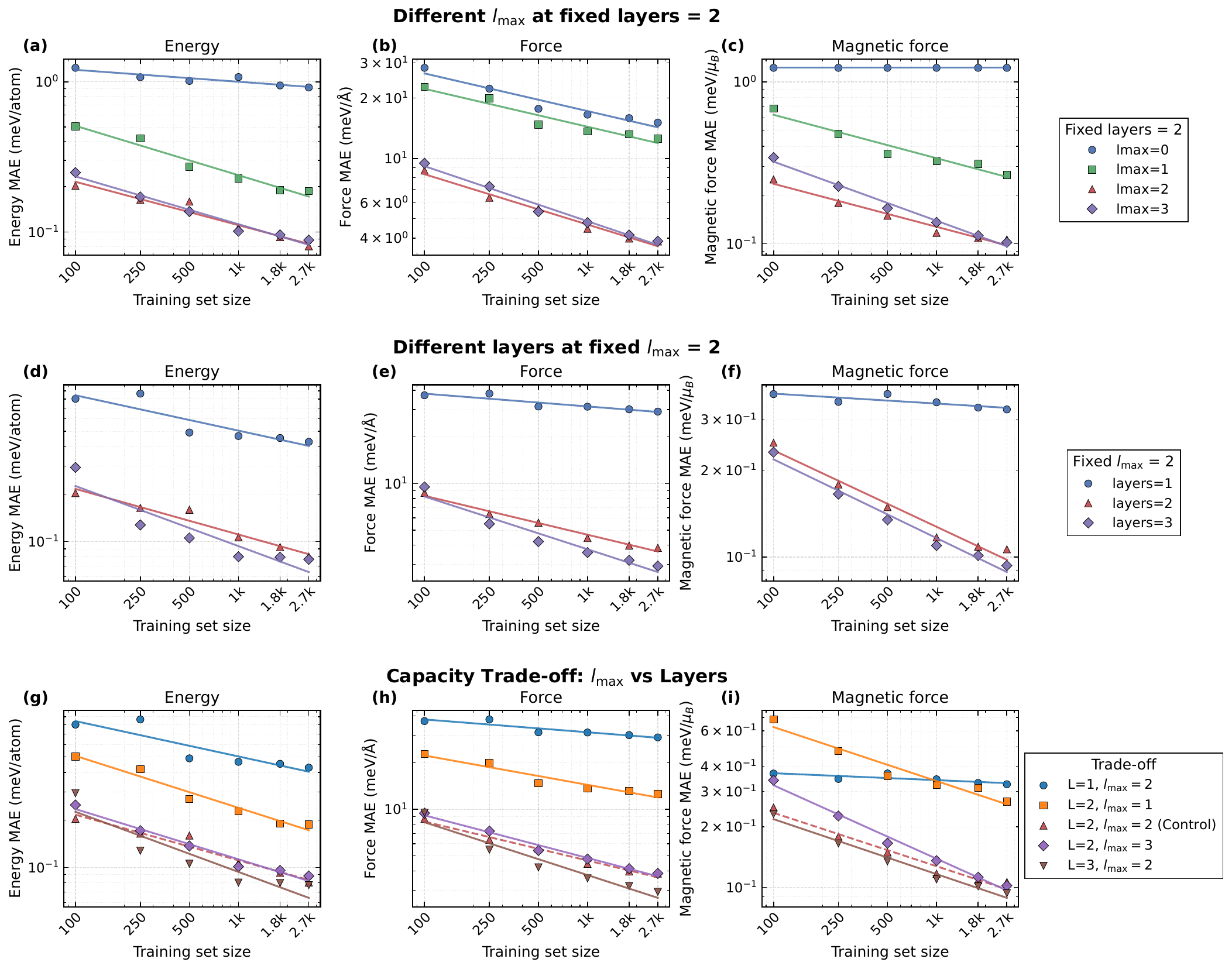}
    \caption{
        Learning curves of STEP on the monolayer CrI$_3$ dataset. 
        (a--c) Scaling curves of energy, force, and magnetic force for different $l_{\max}$ at fixed $L=2$. The scalar model ($l_{\max}=0$) exhibits a failure to resolve magnetic force scaling. 
        (d--f) Scaling behavior across varying layers $L$ at fixed $l_{\max}=2$, showing the benefit of increasing interaction depth. 
        (g--i) Architectural trade-off analysis comparing configurations with broadly similar parameter counts or per-epoch training costs, highlighting the different effects of increasing interaction depth and tensorial order.
    }
    \label{fig:data_efficiency}
\end{figure*}

To evaluate the data efficiency of the STEP model, we investigate its scaling behavior, namely the systematic reduction of the prediction error $\varepsilon$ with the training set size $N$, expressed as $\varepsilon \propto N^{\alpha}$ \cite{scaling_laws,nequip}. Here, $L$ denotes the number of STEP interaction layers, and $l_{\max}$ denotes the maximum angular order retained in the magnetic-moment spherical harmonics and latent equivariant feature channels.

As a representative system exhibiting pronounced anisotropic magnetic interactions and magnetocrystalline anisotropy, with a Curie temperature of approximately 45 K, monolayer CrI$_3$ is a well-known ferromagnetic insulator\cite{dillon1965magnetization,huang2017layer,sarkar2021magnetic}. We constructed a dataset for this material to validate our STEP framework. The full dataset comprises 3290 structures, as detailed in Appendix~\ref{sec:appendix_cri3_dataset}. The dataset was divided into 2700 training, 290 validation, and 300 test configurations. We further utilize a nested dataset splitting strategy on the $\text{CrI}_3$ dataset, constructing training subsets from $N=100$ to $2700$ such that smaller sets are strict subsets of larger ones. The test MAEs for energy, forces, and magnetic forces were evaluated on the same fixed test set of 300 configurations.

The results demonstrate that the magnetic tensor order ($l_{\max}$) is critical for capturing magnetic interactions. With the network depth fixed at $L=2$, a pure scalar network ($l_{\max}=0$) fails to reduce the magnetic force error as more data is provided ($\alpha \approx 0.000$), indicating that invariant scalar representations are insufficient for mapping the complex, orientation-dependent magnetic force landscape \cite{e3nn}. The introduction of higher-order tensors ($l_{\max} \ge 1$) immediately enables systematic error reduction. Specifically, the $l_{\max}=3$ configuration achieves steep scaling exponents of $-0.318$, $-0.275$, and $-0.365$ for energy, force, and magnetic force errors, respectively.

The number of interaction layers $L$ also plays a decisive role in 
determining the learning efficiency, especially for magnetic forces.
The comparison between $L=1,l_{\max}=2$ and $L=2,l_{\max}=1,2$ further 
highlights the importance of the iterative center-environment tensor 
product. Despite having access to higher-order tensor channels, the 
single-layer $L=1,l_{\max}=2$ model exhibits an almost negligible 
magnetic-force scaling exponent ($-0.034$). By contrast, introducing a 
second interaction layer improves the exponent to about $-0.27$, even 
for $l_{\max}=1$. This indicates that the efficient learning of magnetic forces is not governed solely by angular resolution, but also relies critically on repeated coupling between the central spin representation and its spin-lattice environment. Since magnetic forces correspond to gradients of the energy with respect to magnetic degrees of freedom, this behavior provides evidence that the STEP center-environment tensor product facilitates the learning of local magnetic energy gradients. Increasing the depth further to $L=3$ results in even more efficient overall learning, with the force slope reaching $-0.341$ and the magnetic force error remaining consistently low at large training-set sizes.

To examine architectural trade-offs under broadly comparable computational capacity, we compare model configurations selected to have either similar parameter counts or similar per-epoch training costs. For energy and force, increasing the network depth (e.g., $L=2 \to 3$ at $l_{\max}=2$) yields steeper scaling slopes and lower absolute errors than increasing the tensorial order (e.g., $l_{\max}=2 \to 3$ at $L=2$). For magnetic forces, increasing $l_{\max}$ instead produces the steeper learning curve ($-0.365$ versus $-0.273$), whereas the deeper $L=3$ model maintains lower absolute errors over the sampled training-set range. This suggests that while higher angular resolution specifically accelerates the learning of magnetic interactions, iteratively applying the STEP center-environment tensor product (increasing $L$) yields lower absolute errors over the sampled training-set range, remaining a highly effective and robust strategy for minimizing overall prediction errors under constrained computational resources.

\begin{table}[H]
\centering
\caption{Summary of fitted power-law exponents from empirical learning curves for different STEP configurations on the $\text{CrI}_3$ dataset.}
\begin{tabular*}{\linewidth}{@{\extracolsep{\fill}}lccc}
\toprule
Model ($L, l_{\max}$) & Energy & Force & Magnetic force \\
\midrule
$L=1, l_{\max}=2$ & $-0.224$ & $-0.082$ & $-0.034$ \\
$L=2, l_{\max}=0$ & $-0.079$ & $-0.188$ & $0.000$  \\
$L=2, l_{\max}=1$ & $-0.329$ & $-0.189$ & $-0.268$ \\
$L=2, l_{\max}=2$ & $-0.289$ & $-0.251$ & $-0.266$ \\
$L=2, l_{\max}=3$ & $-0.318$ & $-0.275$ & $-0.365$ \\
$L=3, l_{\max}=2$ & $-0.381$ & $-0.341$ & $-0.273$ \\
\bottomrule
\end{tabular*}
\label{tab:slopes}
\end{table}

\begin{figure*}[htbp]
    \centering
    \includegraphics[width=0.9\linewidth]{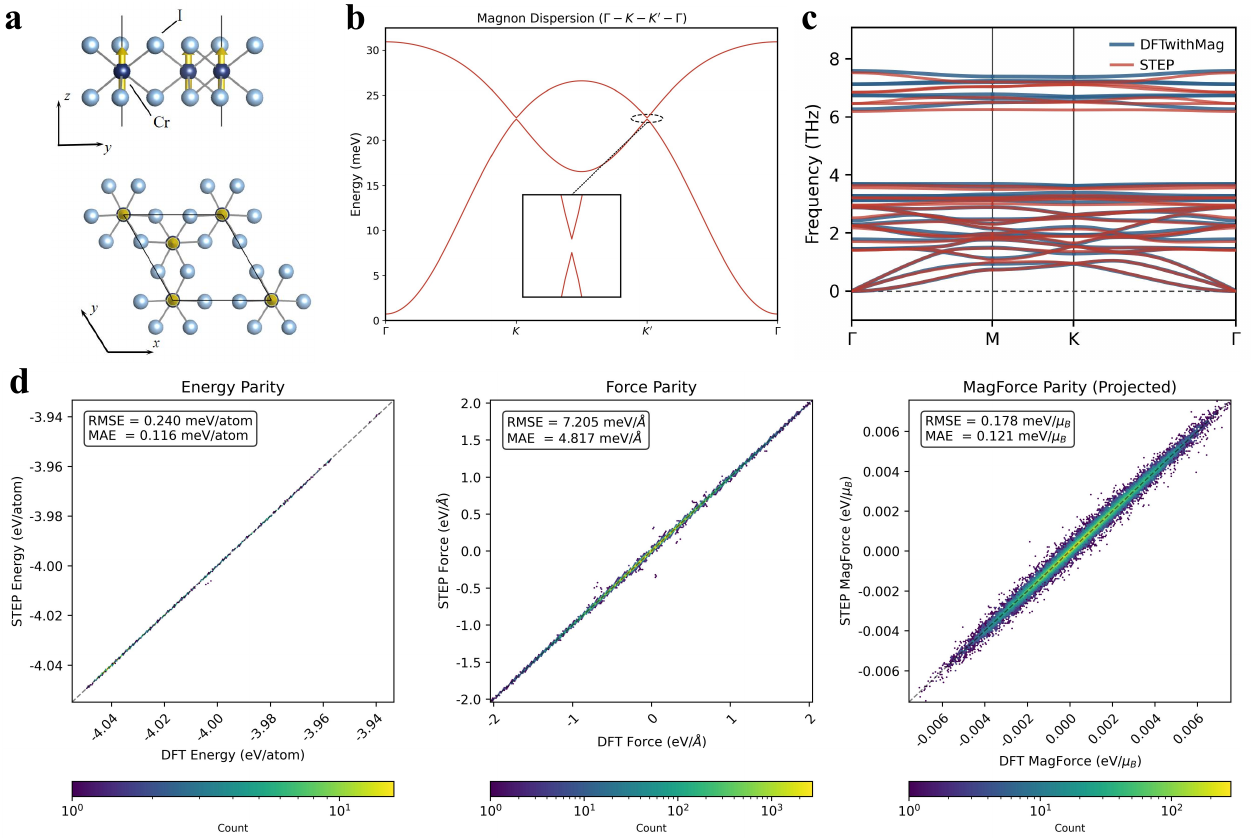}
    \caption{
    Validation of STEP on monolayer CrI$_3$.
    (a) Side and top views of the ferromagnetic ground-state monolayer CrI$_3$ structure. Arrows indicate the local Cr magnetic moments in the ground-state spin configuration, while Cr and I atoms are shown in different colors.
    (b) Magnon dispersion along $\Gamma$--K--K$'$--$\Gamma$ obtained from the learned magnetic energy landscape. Finite gaps are observed at $\Gamma$, K, and K$'$, indicating spin-anisotropic interactions beyond an isotropic Heisenberg description.
    (c) Phonon dispersion along the high-symmetry path $\Gamma$--M--K--$\Gamma$, compared with the spin-polarized DFT reference. STEP reproduces the main acoustic and optical branches with good overall agreement.
    (d) Parity plots over the 900-configuration compact dataset, comparing STEP predictions with DFT reference energies, forces, and projected magnetic forces.
    }
    \label{fig:cri3}
\end{figure*}

\subsection{Compact-data validation on monolayer CrI$_3$}

CrI$_3$ also provides a representative test for evaluating whether STEP can consistently describe both structural and magnetic properties within a unified framework. To evaluate STEP under data-constrained conditions, we trained the model on a compact subset of 900 representative configurations, selected from the full 3290-structure DFT dataset using PCA\cite{pca} and FPS\cite{fps}, as described in Appendix~\ref{sec:appendix_cri3_dataset}. The structural and magnetic coverage of this subset is summarized in Fig.~\ref{fig:dataset_CrI3_900} in Appendix A.

For the training on monolayer CrI$_3$, the STEP model was constructed using two equivariant message-passing layers with a cutoff radius set to $8.0$ \AA. The magnetic-moment spherical harmonics and latent equivariant feature channels were expanded up to $l_{\max}=2$. Edge angular features were used in the message-passing tensor products and projected into the same latent feature space. For the radial embedding, we employed 8 radial basis functions. The radial filter weights were generated by a two-layer MLP with a hidden structure of $8 \rightarrow 64 \rightarrow 64$. The layer-wise energy readout networks each comprised a two-layer MLP ($8 \rightarrow 64 \rightarrow 1$). During training, the dataset was randomly partitioned into training and validation sets with a 9:1 ratio.

As shown in Fig.~\ref{fig:cri3}(d), the predicted energies, forces, and magnetic forces are all in close agreement with the DFT references, indicating that the model captures the local energy landscape with high fidelity for both atomic and spin degrees of freedom. Notably, this predictive accuracy is achieved using only the compact 900-configuration dataset, highlighting the data efficiency of the physically structured STEP representation.
We further examine collective excitations in CrI$_3$. Figure~\ref{fig:cri3}(c) shows the phonon dispersion of ferromagnetic CrI$_3$ along the high-symmetry path $\Gamma$--M--K--$\Gamma$. The STEP results agree well with the DFT phonon spectrum over the full frequency range, reproducing the main acoustic and optical branches along the selected path, which indicates that STEP provides a reliable description of the local curvature of the energy surface with respect to atomic displacements.

We then analyze the magnon spectrum obtained from the learned magnetic energy landscape. As shown in Fig.~\ref{fig:cri3}(b), the magnon dispersion along $\Gamma$--$K$--$K'$--$\Gamma$ exhibits finite gaps at $\Gamma$, $K$, and $K'$. The gap at $\Gamma$ indicates magnetic anisotropy in the effective magnetic Hamiltonian, while the finite gaps at $K$ and $K'$ further show that the learned magnetic Hamiltonian goes beyond purely isotropic Heisenberg exchange interactions, successfully capturing the characteristic magnon-gap features induced by anisotropic exchange couplings \cite{CrI3_magnon}. These results indicate that STEP captures the subtle anisotropic interactions associated with SOC in CrI$_3$.

\subsection{Magnon polarons in Fe$_2$Mo$_3$O$_8$}
\label{sec:femoo_magnon_phonon}

\begin{figure*}[htbp]
    \centering
    \includegraphics[width=\textwidth]{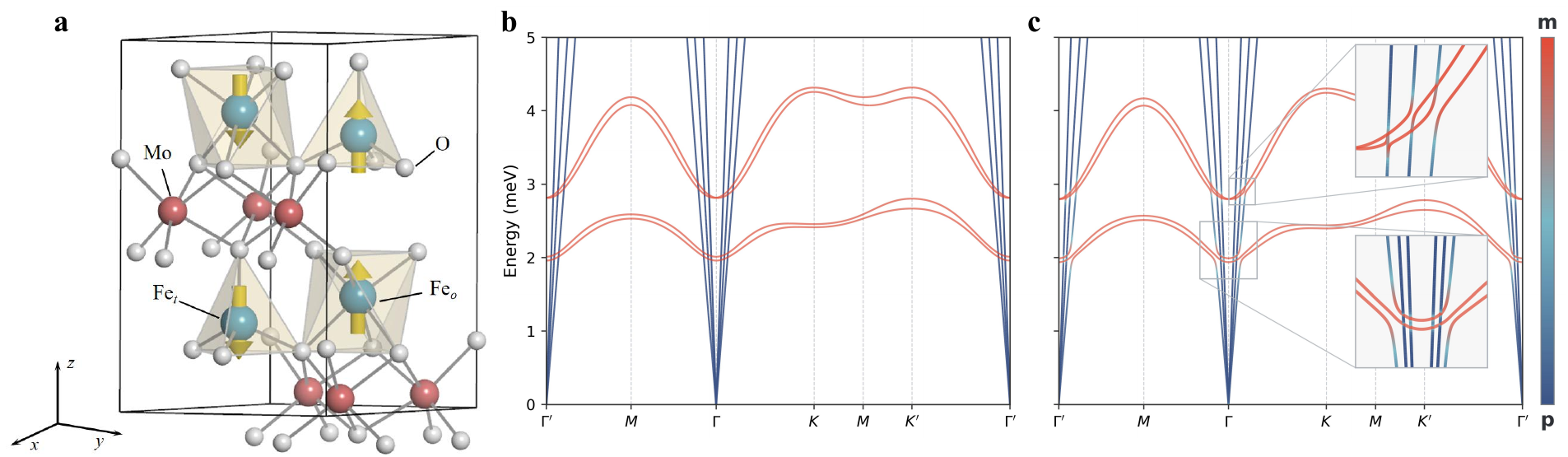}
    \caption{
    Magnon--phonon coupling in Fe$_2$Mo$_3$O$_8$ predicted from the linearized STEP energy surface.
    (a) Crystal and ground state magnetic structure, with inequivalent Fe$_t$ and Fe$_o$ sites indicated.
    (b) Uncoupled magnon and phonon spectra obtained with $K_{u\chi}=0$.
    (c) Hybridized spectrum obtained with the full spin-lattice cross Hessian $K_{u\chi}$, showing avoided crossings characteristic of magnon polaron.
    }
    \label{fig:femoo_magnon_phonon}
\end{figure*}

Fe$_2$Mo$_3$O$_8$ is a polar honeycomb antiferromagnet crystallizing in the P6$_3$mc structure ($a$ = 5.773 {\AA}, $c$ = 10.054 {\AA})\cite{bertrand1975etude,mcalister1983magnetic}. Below $T_N$ $\sim$ 60 K, ferrimagnetic Fe-O layers arising from unequal tetrahedral (Fe$_{t}$ = 4.21 $\mu_B$) and octahedral (Fe$_{o}$= 4.83 $\mu_B$) Fe$^{2+}$ moments stack antiferromagnetically along the $c$ axis\cite{bertrand1975etude}. Its strong spin-lattice coupling has made it a model system for exploring topological magnon polarons\cite{bao2023direct,wu2023fluctuation}: when a magnon branch and a phonon branch approach each other in both energy and momentum, strong hybridization between the two excitations opens a finite gap at the nominal crossing point and produces two separated mixed branches \cite{bao2023direct,wu2023fluctuation}. Reproducing this anticrossing behavior is therefore a direct probe of whether the learned potential captures the strong coupling between atomic displacements and spin fluctuations.

For this analysis, we used the trained STEP energy surface to construct the linearized coupled spin-lattice dynamics around the magnetic ground state of Fe$_2$Mo$_3$O$_8$. Denoting the atomic displacement from equilibrium by $u$ and the transverse spin fluctuation by $\chi$, the energy was expanded to second order as
\begin{equation}
E = E_0 + \frac{1}{2}u^{T}K_{uu}u
        + \frac{1}{2}\chi^{T}K_{\chi\chi}\chi
        + u^{T}K_{u\chi}\chi ,
\label{eq:femoo_coupled_expansion}
\end{equation}
where $E_0$ is the equilibrium energy, $K_{uu}$ is the force-constant matrix, $K_{\chi\chi}$ is the transverse magnetic stiffness matrix, and $K_{u\chi}$ is the spin-lattice cross Hessian coupling atomic displacements to transverse spin fluctuations. Introducing the mass-normalized phonon coordinate $x=M^{1/2}u$, we define
\begin{equation}
D_{\mathrm{ph}} = M^{-1/2}K_{uu}M^{-1/2}, \qquad
G = M^{-1/2}K_{u\chi},
\label{eq:femoo_mass_normalization}
\end{equation}
where $D_{\mathrm{ph}}$ is the phonon dynamical matrix and $G$ is the mass-normalized spin-lattice coupling matrix. The corresponding linearized equations of motion are
\begin{equation}
\ddot{x} = -D_{\mathrm{ph}}x - G\chi,\qquad
\dot{\chi} = \mathcal{J}\left(G^\dagger x + K_{\chi\chi}\chi\right),
\label{eq:femoo_coupled_equations}
\end{equation}
where $\mathcal{J}$ is the local precession matrix for the transverse spin variables. Equivalently, by introducing $v=\dot{x}$, the coupled problem can be written as a unified first-order eigenvalue problem,
\begin{equation}
\frac{d}{dt}
\begin{pmatrix}
x\\ v\\ \chi
\end{pmatrix}
=
\begin{pmatrix}
0 & I & 0\\
-D_{\mathrm{ph}} & 0 & -G\\
\mathcal{J}G^\dagger & 0 & \mathcal{J}K_{\chi\chi}
\end{pmatrix}
\begin{pmatrix}
x\\ v\\ \chi
\end{pmatrix}.
\label{eq:femoo_first_order_matrix}
\end{equation}
Solving the eigenvalue problem associated with Eq.~\eqref{eq:femoo_first_order_matrix} yields the coupled magnon--phonon spectrum. Setting $G=0$ recovers the uncoupled phonon and magnon branches, whereas retaining $G$ yields the hybridized modes. The detailed construction of the transverse spin coordinates and Hessian blocks is given in Appendix~\ref{sec:mp_linear_response}.

Figure~\ref{fig:femoo_magnon_phonon} summarizes the resulting spectrum. In the absence of the cross term, the calculated magnon and phonon branches intersect along the selected momentum path. After the spin-lattice cross Hessian is included, the degeneracy at the crossing is lifted and the two branches repel each other, leaving a finite gap at the original intersection. This behavior is the expected spectral fingerprint of magnon polarons. Importantly, the gap appears without introducing an explicit model Hamiltonian. It follows from the local curvature of the STEP potential energy surface with respect to both lattice and spin coordinates. It is worth noting that our calculated results should be interpreted as a semi-quantitative reproduction of the experimental magnon polaron spectra. To obtain a correct antiferromagnetic ground state, the underlying DFT+$U$ calculations were performed with a large $U=6$ eV, which makes both the phonon and magnon spectra notably softer than experimental values.

The Fe$_2$Mo$_3$O$_8$ dataset used here contains only 97 DFT configurations, as detailed in Appendix~\ref{sec:appendix_femoo_dataset}, and was designed primarily to test whether STEP can learn the spin-lattice energy surface from a limited amount of first-principles data. Over this full 97-configuration dataset, STEP achieves an energy MAE of 1.101 meV/atom, indicating that the sampled local energy landscape is fitted with high accuracy. Within this scope, the emergence of the magnon polaron spectra demonstrates that STEP captures the essential cross response encoded in $K_{u\chi}$. Together with the CrI$_3$ phonon and magnon benchmarks, this result supports the use of STEP as a unified potential for systems where magnetic and lattice degrees of freedom are intrinsically coupled.

\subsection{Finite-temperature magnetic phase transitions}
Another stringent test of a magnetic machine-learning potential is whether it remains reliable not only for static energies and forces, but also for finite-temperature magnetic phase transitions in systems with distinct magnetic characters. Owing to its efficiency and scalability, STEP can be applied to large-scale finite-temperature spin dynamics simulations. Here we consider two representative examples, monolayer CrI$_3$ and bcc Fe. Monolayer CrI$_3$ is a prototypical magnet with local magnetic moments, whereas bcc Fe is a high-$T_c$ itinerant ferromagnet with non-negligible longitudinal spin fluctuations across its magnetic phase transition. The contrast between these two systems therefore provides a stringent test of the extent to which STEP can describe finite-temperature magnetism across qualitatively different systems.

\subsubsection{Monolayer CrI$_3$}

\begin{figure}[htbp]
    \centering
    \includegraphics[width=0.99\columnwidth]{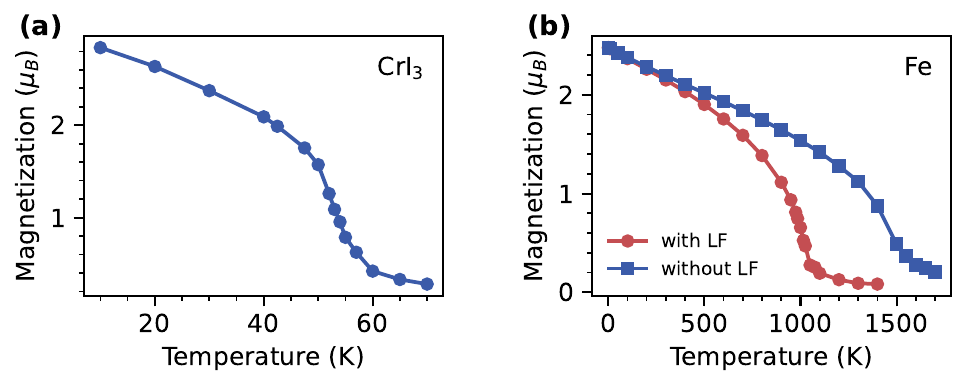}
    \caption{Magnetization per magnetic atom of (a) monolayer CrI$_3$ and (b) bcc Fe. For bcc Fe, results with and without longitudinal fluctuations (LF) are both presented for comparison.} 
    \label{fig:cri3-mag-cv}
\end{figure}

We first examine the finite-temperature magnetic phase transition of monolayer CrI$_3$. Spin dynamics simulations were performed on a $20 \times 20 \times 1$ supercell with in-plane lattice parameters $a=b=6.9685$ \AA\ and $\gamma=120^\circ$. As CrI$_3$ is a magnet with well-defined local magnetic moments, its thermally driven magnetic phase transition is expected to be dominated by transverse spin fluctuations. We therefore fixed the amplitude of each local magnetic moment to be $3~\mu_B$. The simulated temperature dependence of the magnetization is given in Fig.~\ref{fig:cri3-mag-cv}(a), indicating a ferromagnetic to paramagnetic transition at $T_c \sim 55$ K, in good agreement with reported theoretical and experimental results~\cite{Huang2017,Rzepkowski2022}. These results show that STEP captures the essential finite-temperature response of CrI$_3$ as an anisotropic two-dimensional magnet with local magnetic moments.

\subsubsection{bcc Fe}

Since bcc Fe is an itinerant ferromagnet, longitudinal spin fluctuations (LSF) in the magnitudes of local magnetic moments become increasingly important at elevated temperatures, and the rigid spin approximation may consequently overestimate the stability of the ferromagnetic phase~\cite{Ma2012}. To explicitly consider the effect of LSF, we performed generalized Langevin spin dynamics simulations~\cite{Ma2012}. Conventional spin dynamics simulations without LSF were also carried out for comparison. Further details of the spin dynamics formalism and simulation protocol are provided in Appendix~\ref{app.d}. As shown in Fig.~\ref{fig:cri3-mag-cv}(b), compared to the experimental $T_c$ of $1043$ K~\cite{Crangle1971}, simulations with rigid spin obtain a $T_c$ $\sim$1500 K, significantly overestimating the $T_c$ by about 50\%. Including LSF reduces the $T_c$ to $\sim$$1000$ K, in good agreement with the experiments.

\begin{figure}[htbp]
    \centering
    \includegraphics[width=0.48\textwidth]{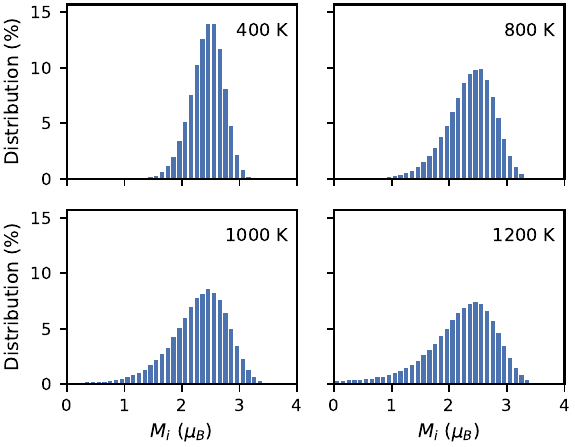}
    \caption{Distribution of local magnetic moments $M_i$ in bcc Fe, shown for $400$, $800$, $1000$, and $1200$ K.}
    \label{fig:fe-moment-distribution}
\end{figure}

To quantitatively show the LSF, we plot the distribution of local magnetic moments at representative temperatures (see Fig.~\ref{fig:fe-moment-distribution}). At $400$ K, the distribution of $M_i$ is narrow and centered around the ground state value, indicating that the system remains in a robust ferromagnetic state with only weak LSF. As the temperature increases to $800$ and $1000$ K, the distribution becomes markedly broader, showing that thermal spin fluctuations in Fe cannot be described solely by rotations of rigid local moments. Instead, longitudinal fluctuations become increasingly important when approaching $T_c$.

Above $T_c$, at $1200$ K, the distribution develops a clear low-moment tail extending toward zero moment. This behavior indicates that the thermal destruction of ferromagnetism in Fe is accompanied not only by orientational disorder, but also by collapse of local moments on a subset of sites. In this sense, the present results directly support the view that longitudinal spin fluctuations are an essential ingredient for a realistic finite-temperature description of itinerant magnets. Taken together, our results show that STEP provides a physically consistent description of itinerant magnetism in bcc Fe.

\section{Conclusions}

We have introduced the Spin Tensor Equivariant Potential (STEP), an equivariant magnetic machine-learning potential that treats vector magnetic moments as continuous geometric degrees of freedom and couples the central magnetic representation to an aggregated local spin-lattice environment through a Center-Environment Tensor Product. Together with equivariant gating, residual interaction updates, and feature-level time-reversal symmetrization, this construction provides a unified energy model from which forces, stresses, and magnetic forces are obtained by differentiation.

Across public FeAl, CrN, and Fe benchmarks, STEP achieves competitive accuracy under the respective evaluation protocols. Learning curves on monolayer CrI$_3$ show that both higher-order tensor channels and iterative center-environment coupling improve the learning of magnetic energy gradients. The resulting potentials reproduce the main phonon and magnon features of monolayer CrI$_3$ and bcc Fe and provide a semi-quantitative description of magnon--phonon hybridization in Fe$_2$Mo$_3$O$_8$. Finite-temperature spin dynamics further yields transition-temperature estimates near 55 K for monolayer CrI$_3$ and 1000 K for bcc Fe, while the comparison between fixed- and variable-length simulations demonstrates the importance of longitudinal moment fluctuations in itinerant Fe. These results establish STEP as a data-efficient and physics-informed framework for coupled spin-lattice energetics, magnetic excitations, and finite-temperature magnetic simulations.

%%%%----------------------------------------
\appendix

\section{Generalized dataset for monolayer CrI$_3$}
\label{sec:appendix_cri3_dataset}

\begin{figure*}[htbp]
    \centering
    \includegraphics[width=0.95\linewidth]{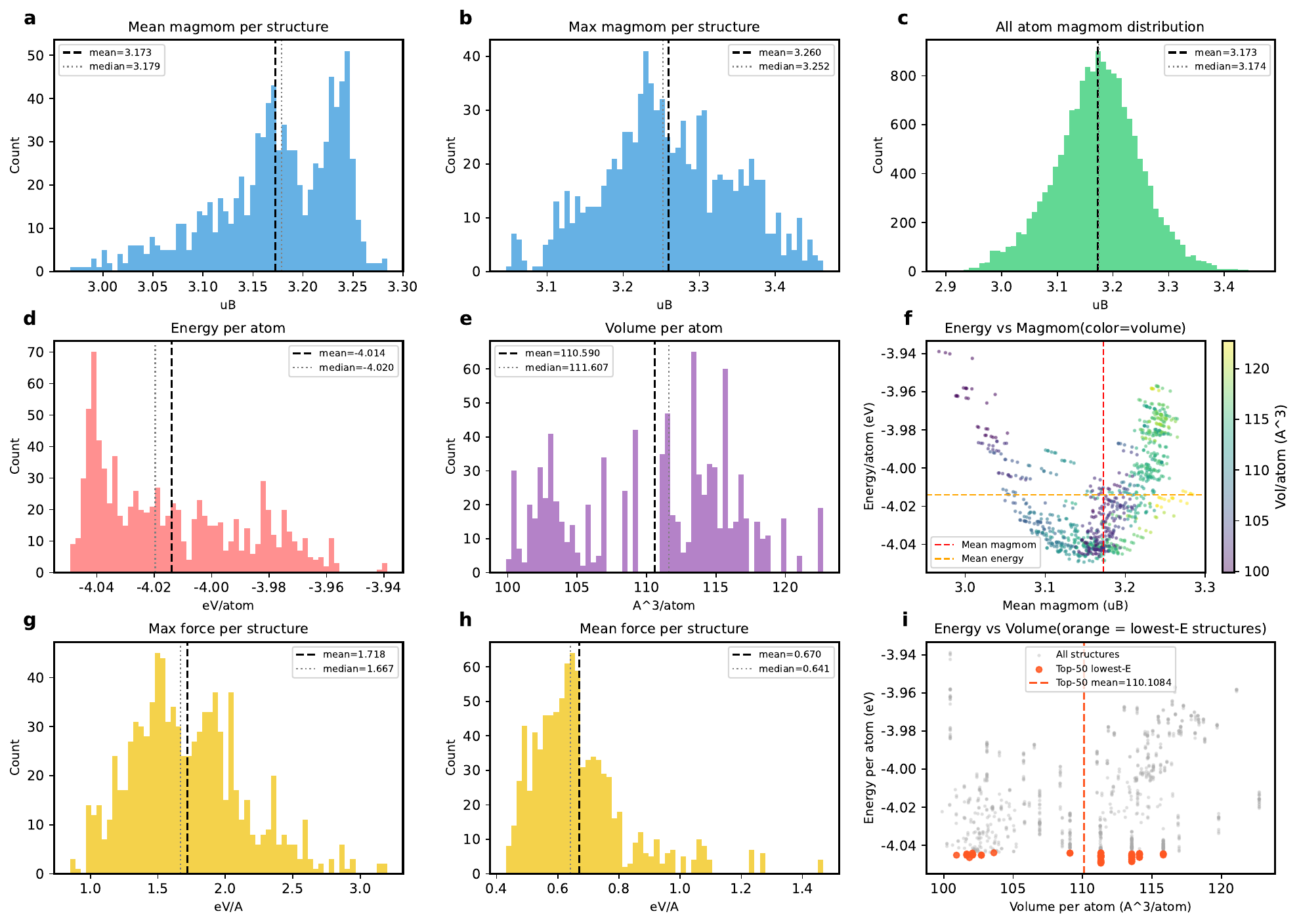}
    \caption{
    \textbf{Distribution of structural and magnetic properties in the CrI$_3$ training dataset.} 
    \textbf{a}--\textbf{c}, Distributions of the mean magnetic moment per structure (\textbf{a}), the maximum magnetic moment per structure (\textbf{b}), and the absolute magnetic moments of all individual magnetic atoms (\textbf{c}), demonstrating a robust coverage of local spin fluctuations around the ferromagnetic ground state.
    \textbf{d},\textbf{e}, Histograms of the potential energy per atom (\textbf{d}) and the volume per atom (\textbf{e}).
    \textbf{f}, Correlation between the energy and the mean magnetic moment, with the point color indicating the volume per atom. The dashed lines represent the dataset averages.
    \textbf{g},\textbf{h}, Distributions of the maximum (\textbf{g}) and mean (\textbf{h}) force within each structure, ensuring sufficient gradient information for force-field parameterization.
    \textbf{i}, Energy--volume phase space of the dataset. The top-50 lowest-energy configurations are highlighted in orange, showing the sampling density near the theoretical equilibrium state.
    }
    \label{fig:dataset_CrI3_900}
\end{figure*}

All density functional theory (DFT) calculations were performed using the Vienna ab initio Simulation Package (VASP)\cite{VASP_cite1,VASP_cite2}. Spin-orbit coupling (SOC) effects were included. The projector-augmented wave (PAW) method\cite{PAW_cite1,PAW_cite2} was employed together with the Perdew-Burke-Ernzerhof (PBE)\cite{PBE} exchange-correlation functional. The important van der Waals interactions in CrI$_3$ were accounted for using the DFT-D3 correction with Becke-Johnson damping\cite{grimme2011effect}. A plane-wave energy cutoff of 500 eV and a 2 $\times$ 2 $\times$ 1 k-point grid were used. A self-consistent effective Hubbard U\cite{dudarev1998electron} value of 1.7 eV\cite{sarkar2021magnetic} was applied to the localized Cr 3d electrons, as determined from linear-response calculations\cite{cococcioni2005linear}.

The training dataset was generated through a combination of structural perturbations and active learning. Starting from the ground state 3 $\times$ 3 $\times$ 1 supercell of CrI$_3$ (72 atoms), we applied in-plane lattice scaling in the range [0.98, 1.02] with an interval of 0.02. A vacuum layer of 19.8 {\AA} was added in the out-of-plane direction to eliminate spurious interactions between periodic images of the two-dimensional material. For each scaling factor, 1000 random atomic displacements with a maximum amplitude of 0.1 {\AA} were generated. The resulting structures were then sparsified and diversified using the farthest-point sampling method implemented in NepTrainKit\cite{chen2025neptrain}, yielding 150 representative structures with combined lattice scaling and atomic perturbations.

We first performed DFT calculations on these 150 perturbed structures in the ferromagnetic spin configuration to obtain energies and related properties, forming the initial dataset. Using the Neuroevolution-potential (NEP) framework\cite{song2024general,xu2025gpumd}, we trained a preliminary model on these structures. Based on this model, we conducted molecular dynamics (MD) simulations on the 3 $\times$ 3 $\times$ 1 CrI$_3$ supercell under non-magnetic conditions. Two thermal protocols were employed: (i) continuous heating from 100 K to 500 K followed by continuous cooling back to 100 K, and (ii) stepwise heating and cooling with each 100 K interval held for 1 ns. In both cases, the out-of-plane box dimension was fixed at 19.8 {\AA} throughout the simulations. From these trajectories, 320 structures were collected and subjected to DFT calculations in the ferromagnetic configuration.

For all 470 structures obtained so far (150 perturbed + 320 from MD), we generated additional spin configurations as follows: for each structure, two random spin configurations were created for every Cr atom with a maximum deviation of 90$^\circ$ from the primary axis, and four completely random magnetic orders were also produced. DFT calculations with the same accuracy were performed on all these spin-configured structures. This process ultimately produced the complete dataset of 3290 structures encompassing a wide variety of spin configurations.

For the compact 900-configuration subset used in the main CrI$_3$ validation, the distributions of magnetic moments, energies, volumes, and force magnitudes are summarized in Fig.~\ref{fig:dataset_CrI3_900}, confirming that the selected subset retains broad structural and magnetic coverage.

As illustrated in the data distribution profiles (Fig.~\ref{fig:dataset_CrI3_900}), this compact dataset effectively encompasses a broad structural and magnetic phase space. Specifically, the potential energy spans from $-4.05$ to $-3.94$ eV/atom, and the atomic volume varies significantly from $99.88$ to $122.73$ \AA$^3$/atom, capturing the essential energy variations associated with lattice expansion, compression, and other structural changes. Furthermore, the dataset contains diverse spin-lattice coupling environments: the magnitude of the atomic magnetic moments fluctuates around an average of $3.17\ \mu_B$, with individual atomic moments ranging from $2.89$ to $3.46\ \mu_B$. The structures also exhibit a robust distribution of forces (mean value $\sim 0.67$ eV/\AA, maximum up to $3.20$ eV/\AA), providing sufficient gradient information for accurate force-field parameterization.

\section{Dataset for Fe$_2$Mo$_3$O$_8$}
\label{sec:appendix_femoo_dataset}

Consistent with the calculations performed on CrI$_3$, DFT single-point calculations on Fe$_2$Mo$_3$O$_8$ were carried out using VASP\cite{VASP_cite1,VASP_cite2} with the PAW method\cite{PAW_cite1,PAW_cite2} and PBE exchange-correlation functional\cite{PBE}. SOC effects were included self-consistently. To correctly stabilize the antiferromagnetic (AFM) ground state, the on-site Coulomb interaction U and exchange interaction J for the Fe 3d orbitals were treated using the fully rotationally invariant LSDA+U approach\cite{dudarev1998electron} introduced by Liechtenstein et al.\cite{Liechtenstein}, with U = 6 eV and J = 0 eV. Owing to the default Wigner-Seitz sphere integration volume in the pseudopotentials, the calculated magnetic moments in Fe$_2$Mo$_3$O$_8$ are smaller than the experimental values (Fe$_{t}$ = 3.78 $\mu_B$ and Fe$_{o}$ = 3.81 $\mu_B$ in our calculations). This underestimation has also been noted in previous studies by Chang et al.\cite{Chang}.

The dataset was constructed as follows. Experimental lattice constants were employed, with two types of supercells ( 2$\times$1$\times$1 and a 1$\times$1$\times$2) generated to balance computational cost and magnetic configuration sampling. For each type of supercell, 50 structures with random atomic displacements (maximum amplitude 0.1 {\AA}) were created. Each structure was initialized with a fully random orientation of the magnetic moments. Single-point DFT calculations were performed on the resulting 100 structures. After excluding 4 structures that failed to converge, and including the ground-state AFM configuration, a total of 97 structures were collected as the final dataset.

\section{Magnon--phonon spectra from the STEP energy surface}
\label{sec:mp_linear_response}

The coupled magnon--phonon spectrum was obtained by linearizing the learned STEP energy surface around the relaxed antiferromagnetic reference configuration. Let $\mathbf{R}_0$ and $\{\mathbf{S}_i^0\}$ denote the equilibrium atomic positions and magnetic moments. Atomic displacements are defined as $u=\mathbf{R}-\mathbf{R}_0$. For each magnetic atom, the spin fluctuation is projected onto the local tangent plane perpendicular to $\mathbf{S}_i^0$,
\begin{equation}
\delta\mathbf{S}_i =
\chi_{i1}\mathbf{e}_{i1}
+\chi_{i2}\mathbf{e}_{i2},
\qquad
\mathbf{e}_{ia}\cdot\mathbf{S}_i^0=0 ,
\label{eq:appendix_transverse_spin}
\end{equation}
where $\{\mathbf{e}_{i1},\mathbf{e}_{i2}\}$ is a local orthonormal transverse basis and $\chi$ collects all transverse spin coordinates. Keeping only the harmonic terms gives
\begin{equation}
E = E_0 + \frac{1}{2}u^{T}K_{uu}u
        + \frac{1}{2}\chi^{T}K_{\chi\chi}\chi
        + u^{T}K_{u\chi}\chi ,
\label{eq:appendix_harmonic_energy}
\end{equation}
with
\begin{equation}
K_{uu}=\frac{\partial^2E}{\partial u\,\partial u},\quad
K_{\chi\chi}=\frac{\partial^2E}{\partial \chi\,\partial \chi},\quad
K_{u\chi}=\frac{\partial^2E}{\partial u\,\partial \chi},
\label{eq:appendix_hessian_definitions}
\end{equation}
evaluated at $(\mathbf{R}_0,\{\mathbf{S}_i^0\})$. The three Hessian blocks were evaluated from the differentiable STEP energy surface, so that the force-constant, magnetic-stiffness, and spin-lattice coupling matrices are mutually consistent derivatives of the same learned potential energy surface.

The lattice part is converted to the standard phonon form by the mass-normalized coordinate $x=M^{1/2}u$:
\begin{equation}
D_{\mathrm{ph}}=M^{-1/2}K_{uu}M^{-1/2},
\qquad
G=M^{-1/2}K_{u\chi}.
\label{eq:appendix_mass_normalized}
\end{equation}
The linearized Newton equation for the lattice variables is therefore
\begin{equation}
\ddot{x}=-D_{\mathrm{ph}}x-G\chi .
\label{eq:appendix_phonon_equation}
\end{equation}
The transverse spin dynamics is written in first-order precessional form,
\begin{equation}
\dot{\chi}=\mathcal{J}\left(G^\dagger x+K_{\chi\chi}\chi\right),
\label{eq:appendix_spin_equation}
\end{equation}
where $\mathcal{J}$ is a block-diagonal antisymmetric matrix determined by the local spin precession convention. Combining Eqs.~\eqref{eq:appendix_phonon_equation} and \eqref{eq:appendix_spin_equation} with $v=\dot{x}$ gives
\begin{equation}
\frac{d}{dt}
\begin{pmatrix}
x\\ v\\ \chi
\end{pmatrix}
=
\begin{pmatrix}
0 & I & 0\\
-D_{\mathrm{ph}} & 0 & -G\\
\mathcal{J}G^\dagger & 0 & \mathcal{J}K_{\chi\chi}
\end{pmatrix}
\begin{pmatrix}
x\\ v\\ \chi
\end{pmatrix}.
\label{eq:appendix_first_order_matrix}
\end{equation}
For a normal mode with time dependence $\exp(i\omega t)$, Eq.~\eqref{eq:appendix_first_order_matrix} becomes a matrix eigenvalue problem with eigenvalues $\lambda=i\omega$. The positive-frequency branches extracted from these eigenvalues give the coupled magnon--phonon spectrum. In the calculations shown in Fig.~\ref{fig:femoo_magnon_phonon}, the uncoupled reference spectrum was obtained by setting $G=0$, while the hybridized spectrum was obtained by retaining the full spin-lattice cross Hessian $G$.

\section{Details of spin molecular dynamics simulations}
\label{app.d}

Spin-molecular dynamics simulations were performed using the semi-implicit B (SIB) integration scheme implemented in LAMMPS~\cite{Mentink2010, Thompson2022LAMMPS}. In all simulations, the lattice degrees of freedom were kept fixed, and the spin-dynamics time step was set to $1.0$ fs.

\subsection{CrI$_3$}

For monolayer CrI$_3$, a $20\times20\times1$ supercell containing 3200 atoms was employed. The in-plane lattice constant was $6.9685$~\text{\AA}. The simulation cell length along the out-of-plane direction was $20.7559$~\text{\AA}, providing vacuum separation between periodic images of the monolayer. Rigid spin dynamics was employed, with the magnitude of each local magnetic moment set to $3~\mu_B$. The spin system was coupled to a transverse Langevin thermostat with a damping parameter of $0.1$. At each temperature, the system was equilibrated for $20$ ps and subsequently sampled for $100$ ps. Near the magnetic phase-transition region, the sampling time was extended to $200$ ps to improve statistical accuracy.

\subsection{bcc Fe}

For bcc Fe, a $10\times10\times10$ supercell constructed from the conventional cubic unit cell was employed, corresponding to a cubic simulation cell with a side length of $28.6304$~\text{\AA}. Longitudinal fluctuations of local magnetic moments were included using the generalized Langevin spin-dynamics formalism. The local magnetic moment is decomposed as
\begin{equation}
\mathbf{m}_i=m_i\hat{\mathbf{s}}_i,
\qquad
|\hat{\mathbf{s}}_i|=1,
\end{equation}
where $m_i$ and $\hat{\mathbf{s}}_i$ denote the moment magnitude and direction, respectively. The effective magnetic field is calculated as
\begin{equation}
\mathbf{H}_i^{\mathrm{eff}}
=
-\frac{\partial E}{\partial \mathbf{m}_i}.
\end{equation}

The generalized Langevin dynamics combines stochastic transverse spin dynamics with longitudinal relaxation of the moment magnitude. The transverse dynamics follows the stochastic Landau--Lifshitz equation integrated using the SIB scheme proposed by Mentink \textit{et al.},\cite{Mentink2010} whereas the longitudinal dynamics follows the variable-length Langevin spin formalism introduced by Ma and Dudarev.\cite{Ma2012}

The transverse spin dynamics is given by
\begin{equation}
\begin{split}
\frac{d\hat{\mathbf{s}}_i}{dt}
=
-\frac{\gamma}{1+\alpha_t^2}
\Big[
&\hat{\mathbf{s}}_i
\times
\left(
\mathbf{H}_{i,\perp}^{\mathrm{eff}}
+
\boldsymbol{\eta}_{i,T}
\right)
\\
&+
\alpha_t\hat{\mathbf{s}}_i
\times
\left[
\hat{\mathbf{s}}_i
\times
\left(
\mathbf{H}_{i,\perp}^{\mathrm{eff}}
+
\boldsymbol{\eta}_{i,T}
\right)
\right]
\Big],
\end{split}
\label{eq:transverse_glangevin}
\end{equation}
where
\begin{equation}
\mathbf{H}_{i,\perp}^{\mathrm{eff}}
=
\mathbf{H}_i^{\mathrm{eff}}
-
\left(
\mathbf{H}_i^{\mathrm{eff}}
\cdot
\hat{\mathbf{s}}_i
\right)
\hat{\mathbf{s}}_i.
\end{equation}
Here, $\gamma$ is the gyromagnetic ratio, $\alpha_t$ is the transverse damping coefficient, and $\boldsymbol{\eta}_{i,T}$ denotes the transverse thermal noise. The longitudinal dynamics of the moment magnitude is described by
\begin{equation}
\frac{dm_i}{dt}
=
\gamma_L H_{i,\parallel}^{\mathrm{eff}}
+
\xi_{i,L},
\label{eq:longitudinal_glangevin}
\end{equation}
where
\begin{equation}
H_{i,\parallel}^{\mathrm{eff}}
=
\mathbf{H}_i^{\mathrm{eff}}
\cdot
\hat{\mathbf{s}}_i.
\end{equation}

\begin{acknowledgments}

This work was supported by the National Key R\&D Program of China (Grant No.~2023YFB4603801), National Natural Science Foundation of China (Grant Nos.~12474249,U25A20193),
Guangdong Provincial Key Laboratory of Magnetoelectric Physics and Devices (No.~2022B1212010008).

\end{acknowledgments}

\bibliography{STEPref} 

\end{document}